# The Hayabusa Spacecraft Asteroid Multi-Band Imaging Camera: AMICA

Masateru Ishiguro [1]
Ryosuke Nakamura [2]
David J. Tholen [3]
Naru Hirata [4], Hirohide Demura [4] Etsuko Nemoto [4]
Akiko M. Nakamura [5]
Yuta Higuchi [6]
Akito Sogame [7]
Aya Yamamoto [8]
Kohei Kitazato [4], Yasuhiro Yokota [9], Takashi Kubota [9], Tatsuaki Hashimoto [9], and Jun Saito [7]

[1] Department of Physics and Astronomy, Seoul National University, Seoul 151-747, Korea
E-mail: ishiguro@astro.snu.ac.kr
[2] National Institute of Advanced Industrial Science and Technology, Tsukuba 305-8568, Japan
[3] Institute for Astronomy, University of Hawaii, Honolulu, HI 96822, U.S.A
[4] Dept. Computer Software, University of Aizu, Aizu-wakamatsu, Fukushima 965-8580, Japan
[5] Graduate School of Science, Kobe University, Kobe, 657-8501, Japan
[6] Lunar and Planetary Laboratory, University of Arizona, Tucson, AZ 86705-6643, USA
[7] School of Engineering, Tokai University, Hiratsuka, Kanagawa 259-1292, Japan
[8] Remote Sensing Technology Center of Japan (RESTEC), Tokyo 106-0032, Japan
[9] Institute of Space and Astronautical Science (ISAS), Japan Aerospace Exploration Agency (JAXA), Sagamihara, Kanagawa 229-8510, Japan

Offprint Request: ishiguro@astro.snu.ac.kr

**Proposed Running Head**: The Hayabusa Mission Multi-band Imaging Camera: AMICA

**Editorial Correspondence to:**
Masateru Ishiguro, Dr.
Astronomy Department, School of Physics & Astronomy,
College of Natural Sciences, Seoul National University,
Gwanak-gu, Seoul 151-747, Korea
E-mail: ishiguro@astro.snu.ac.kr,
Tel: +82-2-880-4232, Fax: +82-2-887-1435

**ABSTRACT**

The Hayabusa Spacecraft Asteroid Multiband Imaging Camera (AMICA) has acquired more than 1400 multispectral and high-resolution images of its target asteroid, 25143 Itokawa, since late August 2005. In this paper, we summarize the design and performance of AMICA. In addition, we describe the calibration methods, assumptions, and models, based on measurements. Major calibration steps include corrections for linearity and modeling and subtraction of bias, dark current, read-out smear, and pixel-to-pixel responsivity variations. AMICA v-band data were calibrated to radiance using in-flight stellar observations. The other band data were calibrated to reflectance by comparing them to ground-based observations to avoid the uncertainty of the solar irradiation in those bands. We found that the AMICA signal was linear with respect to the input signal to an accuracy of << 1% when the signal level was < 3800 DN. We verified that the absolute radiance calibration of the AMICA v-band (0.55 μm) was accurate to 4% or less, the accuracy of the disk-integrated spectra with respect to the AMICA v-band was about 1%, and the pixel-to-pixel responsivity (flatfield) variation was 3% or less. The uncertainty in background zero-level was 5 DN. From wide-band observations of star clusters, we found that the AMICA optics have an effective focal length of 120.80 ± 0.03 mm, yielding a field-of-view (FOV) of 5.83° x 5.69°. The resulting geometric distortion model was accurate to within a third of a pixel. We demonstrated an image-restoration technique using the point-spread functions of stars, and confirmed that the technique functions well in all loss-less images. An artifact not corrected by this calibration is scattered light associated with bright disks in the FOV.

Keywords: Instrumentation; Asteroid Itokawa; Near-Earth Objects

## 1. Introduction

Hayabusa (formerly known as MUSES-C) is a Japanese space mission designed to collect a sample of surface material from the near-Earth asteroid 25143 Itokawa (1998 SF36) and return the sample to Earth. The spacecraft was launched on 9 May 2003 and arrived at Itokawa on 12 September 2005. After arrival, remote sensing observations were carried out for 2 months to select landing sites and study the asteroid itself (Saito et al., 2006; Fujiwara et al., 2006). In November 2005, Hayabusa landed on Itokawa to collect samples (Yano et al., 2006). It is thought that fine particles were swirled into the sampling capsule (Fujiwara et al. 2006). Hayabusa will return to Earth in June 2010.

During the rendezvous phase with Itokawa, Hayabusa was designed to find its position relative to the asteroid autonomously using optical navigation cameras (ONC) and range data based on light-detection and ranging (LIDAR) technology (Hashimoto et al., 2003; Kubota et al. 2005). The ONC-T, a refractive telescope camera head, is referred to as the Asteroid Multiband Imaging Camera (AMICA) when used for scientific observations. Prior to "touch-and-go sampling," Itokawa was mapped by AMICA to select sampling sites free of serious obstacles such as meter-sized boulders and craters (Hirata et al. 2009; Michikami et al. 2009). Assessing the performance and determining the calibration of AMICA are pivotal to the analysis of these data.

The main goal of this study is to derive the flux calibration parameters for the AMICA in-flight data. We describe the design of AMICA flight model (hereafter AMICA) in Section 2 and the in-flight operations in Section 3. The performance and the calibration method are detailed in Section 4. We compared the calibrated AMICA flux data with those of the ground-based observations. We summarize the paper in Section 5.

**Note and Limitation**

The current AMICA science team was reconstructed in April 2002, about one year before the launch of Hayabusa spacecraft, taking over the former AMICA team. Note that preflight information about the transmittance of the optics was obtained on the basis of documents and e-mails received by the handover process. Since the measurement accuracy of the filter's transmittance and the other optical system was not examined, we cannot disclose the digitized data for the avoidance of misunderstanding. The AMICA data user should not use the transmittance in Figure 3 for the radiometric calibration. Instead, we recommend using the conversion factors determined by the in-flight operations (see Table 9). In addition, we do not deal here with wide-band because it was installed for the engineering purpose (not for the scientific purpose).

## 2. AMICA Design

AMICA is one of the optical navigation cameras (ONC) comprised of three units: three charge-coupled device (CCD) camera heads (ONC-T, ONC-W1, and ONC-W2), an analog signal processing unit (ONC-AE), and a digital signal processing unit (ONC-E). AMICA (ONC-T) and two wide-angle navigation cameras (ONC-W1 and ONC-W2) are controlled through ONC-E and ONC-AE, and the signals obtained are converted from analog to digital in ONC-AE. The digitized images are numerically processed by the ONC-E. The design of the AMICA camera head flight model is essentially the same as that of its prototype model (Nakamura et al., 2002). It is a CCD camera with a refractor telescope consisting of cosmic radiation-resistant and antireflection-coated lenses. The CCD device was configured in a frame-transfer architecture with an image area and a frame storage area. The depth of field (the distance between the camera and objects that are observed in focus) was designed to be 75 m - $\infty$. AMICA was installed in the bottom panel of the spacecraft, with its optical axis boresighted with those of the near-infrared spectrometer (NIRS, Abe et al. 2006a, 2007) and the LIDAR (Mukai et al., 2002, 2006; Abe et al., 2006b) (Fig. 1). A cylindrical stray light baffle, attached in front of AMICA, reduced light from sources outside the field-of-view (FOV). Because AMICA was installed close to the asteroid-sampling device, it was essential to obtain calibration data before touchdown, to avoid the significant degradation by contaminants associated with surface sampling. Table 1 summarizes the AMICA specifications.

[Figure 1]

[Table 1]

The effective FOV was 5.83°×5.69°, which is covered by 1024×1000 pixels, so that the instantaneous FOV (IFOV) corresponded to 20.490"/pixel. This pixel resolution corresponds to ~70 cm from Itokawa's surface at home position (nominal distance of 7 km from Itokawa's surface). The data for each image were stored in a 1024x1024 array, with two 12-column-wide masked areas on the right and left edges of the array to monitor the zero-signal level (Fig. 2). The adopted CCD was a back-illuminated Multi-Pinned Phase (MPP)-type. This type of CCD has two main advantages: the dark current is lower than that of non-MPP type CCDs, and it has a highly uniform subpixel scale response. The inflight temperature of the CCD (between -18 and -30ºC) was low enough such that there was no appreciable dark as a function of integration time. The subpixel response uniformity made geometric and photometric calibrations from stellar observations simpler (see subpixel nonuniformities of NEAR/MSI, Murchie et al., 1999).

[Figure 2]

The exposure time was controlled electronically. AMICA has no mechanical shutter. We can designate any of 30 different exposure times, ranging from 5.44 ms to 178 s, plus a zero-second exposure (more precisely, less than 1 μs). The wide exposure time is required to image both the bright asteroid and the dim stars.

The ONC-E enables numerical operations from data in up to 16-frame buffers from multiple images in arbitrary combinations. In a nominal imaging sequence, we took a pair of images in a short time interval (typically 15 s) and then used the ONC-E to subtract the latter frame (smear frame, 0-s exposure) from the former frame (≥ 5.44-ms exposure). This procedure simultaneously corrected for smear, dark current, and bias. We assumed the dark current was constant during the time interval between the two exposures. To reduce the total amount of data downlinked to Earth stations, we occasionally performed binning, extraction of partial areas of interest, and reversible or irreversible data compressions before sending the data to the onboard data recorder from the ONC-E buffer. Binning mode generated the 'average' of pixel values within the boxed area.

AMICA has an eight-filter system, seven of which are compatible with the Eight Color Asteroid Survey (ECAS), the standard for ground-based asteroid taxonomy (Zellner et al., 1985). The ECAS comprehensive database is available via the Internet on the NASA Planetary Data System (PDS) server (http://pds.jpl.nasa.gov). In addition, a neutral density (ND) filter was attached because the target was changed from a low-albedo asteroid to a high-albedo asteroid. The original mission target was 4660 Nereus (whose albedo was considered to be 0.05, Ishibashi et al. 2000). Due to scheduling delays, the mission target was changed twice, to 1989 ML and then to Itokawa (Yoshikawa et al. 2000). The large discrepancy in albedo between Nereus and Itokawa caused us modify the optical transmittance. The transmittance of the ND filter is 3.7% at 550 nm. The transmittance of each of the seven filters, the ND filter, and the other optics were measured in a pre-flight standalone experiment. The effective wavelengths of AMICA's seven band filters were computed as follows:

$$\lambda_{\text{eff}} = \frac{\int \lambda\phi(\lambda)S(\lambda)d\lambda}{\int \phi(\lambda)d\lambda \int S(\lambda)d\lambda} \quad , \tag{1}$$

where $\phi(\lambda) = T_{\text{FIL}}(\lambda)$ x $T_{\text{ND4\%}}(\lambda)$ x $T_{\text{OPT}}(\lambda)$ x $QE(\lambda)$ is the system spectral efficiency; $T_{\text{FIL}}(\lambda)$, $T_{\text{ND4\%}}(\lambda)$ and $T_{\text{OPT}}(\lambda)$ are the transmissions of each of the seven filters, the ND filter, and the other optical systems, respectively, measured in pre-flight standalone experiments; $QE(\lambda)$ is the nominal quantum efficiency of the CCD provided by the manufacturer; and $S(\lambda)$ represents the spectrum of the light source. The effective wavelengths were calculated for sunlight, and the result is shown in Table 2. AMICA's effective wavelengths were within ~10 nm of standard ECAS in the b-, v-, w-, x- and p-bands, but diverged from the standard ECAS (Tedesco et al, 1982) in the ul- and zs-bands. An additional wide-band (clear) filter was installed in the filter wheel for faint-object detection (e.g., satellite survey, Fuse et al., 2007), and four polarizers were attached to the upper left edge of the CCD chip (Fig. 2). Figure 3 provides the AMICA filter transmittance and the system spectral efficiency. The ul-filter exhibited a weak leak at longer wavelengths (> 830 nm) and the intensity contribution of solar-colored objects was estimated to increase the signal by 1.4% over the signal in the designed wavelength (310–410 nm). Table 2 compares the effective wavelengths and widths of the AMICA filters to those of the ECAS standard system (Tedesco et al, 1982).

[Figure 3]

[Table 2]

Figure 4 shows a front view of AMICA. The large cylinder at the center of the image is the filter wheel casing, and the small cylinder at the bottom is the entrance to the optics. The two protrusions located in front of the optics are small lamps that irradiate the optics. They are referred to as FF-lamps, and are used for degradation monitoring in sensitivity degradation (see Section 4.2.2).

[Figure 4]

## 3. AMICA Operation Summary

The in-flight operation phases are conventionally referred to as the Initial Operation Phase (IOP), Cruise Phase (CP), Approach Phase (AP), and Mission Phase (MP). In this section, we summarize the AMICA operations during each phase.

During the IOP, instrument functioning was tested. Following launch on 9 May 2003, we checked the device functioning. AMICA took FF-lamp images to investigate whether any contamination had adhered to the camera. On 23 May 2003, we took stellar images of the area around the head of Scorpius.

The CP was the period during which the spacecraft was driven by the ion engine. We observed stars during breaks in ion engine operation. $\tau$ Sco was imaged again on 1 June 2004 to monitor the degradation in sensitivity because $\tau$ Sco is not likely to be a variable star. The Hayabusa spacecraft encountered Earth on 19 May 2004, receiving a gravity assist toward

Itokawa's orbit. During the Earth swing-by, we took a sequence of Earth and Moon images. On 26 April 2005, we imaged a portion of sky in Taurus with the wide-band filter for geometric calibration. Unfortunately, we could not obtain the images with the other seven filters for the geometric calibration because of the limited observation time.

During the AP, the spacecraft approached Itokawa using chemical engines, approaching to within ~10,000 km of Itokawa. On 22 August 2005, we obtained the first image of Itokawa from AMICA. After that, we observed brightness variations in Itokawa for comparison with ground-based lightcurves. On 29 August, $\alpha$ Leo was included in images of Itokawa. On 1 September, Itokawa's Hill Sphere was surveyed for possible satellites through the wide-band filter (Fuse et al., 2007). The Hayabusa spacecraft arrived at Itokawa at 01:00 UTC, 12 September 2005.

During the MP, the spacecraft did not orbit Itokawa, but remained in a station-keeping heliocentric orbit. Hayabusa initially observed the asteroid's surface from a distance of ~20 km in the Gate Position (GP), a region roughly on a line connecting Earth to the asteroid on the sunward side. Global color mapping (Ishiguro et al., 2007) and shape modeling (Demura et al., 2006; Gaskell et al., 2008) were also performed. On 27 September, a polarimetric observation was conducted at a 10° phase angle. Higher-resolution imaging was performed from a distance of about 7 km (Home Position, hereafter HP). AMICA covered the whole surface of Itokawa in both the GP and HP. On 3 October 2005, Hayabusa lost the use of the $Y_{SC}$-axis reaction wheel (see Fig. 1), and thereafter used one reaction wheel and two chemical thrusters to maintain attitude control. This malfunction made it difficult to obtain polarimetric data at larger solar phase angles, and the Hayabusa project team decided to cancel the high solar phase angle observations (~90°). These were replaced by Tour observations, in which the spacecraft moved to various attitudes and solar phase angles (Sun-Itokawa-AMICA angle, <35°) to access the polar region and make high-resolution topographic images under different lighting conditions (Fujiwara et al., 2006). Using the observations in the GP, HP, and Tour sub-phases, a surface sampling location on a smooth area called Muses-C Regio was selected.

During the MP, the spacecraft moved close to the surface on 4 and 12 November for touchdown rehearsals. First and second touchdowns were made on 19 and 25 November. During the descending and ascending periods, AMICA took close-up images of Itokawa (Yano et al. 2006; Miyamoto et al. 2007; Nakamura et al. 2007; Noguchi et al. 2009). In addition, AMICA obtained a set of stellar images and FF-lamp images for the purpose of the calibration.

## 4. Inflight calibrations

AMICA raw data (Level 1) are archived on the Internet servers at ISAS (http://darts.isas.jaxa.jp/planet/hayabusa/), and will be archived by the NASA/Planetary Data System (PDS, http://pds.jpl.nasa.gov/). All calibration data obtained during in-flight operation are summarized in Table 3. In this section, we review AMICA performance based on these data sets.

[Table 3]

### 4-1. Geometric calibrations

#### 4-1-1. Co-alignment with NIRS and LIDAR

The optical axis of AMICA was designed to be boresighted with NIRS and LIDAR. Accurate knowledge of the NIRS and LIDAR pointing directions in AMICA coordinates was required to extend the NIRS spectral range to the visible wavelength using the AMICA filter set, and improve the measurement accuracy of the distance between the spacecraft and the LIDAR footprint. The AMICA NIR bands were essential to characterize the mafic absorption band around 1 μm because in conjunction with the NIRS (Abe et al., 2005, 2006a). Since the w-band was designed to match the peak of the continuum for an S-type asteroids (Tedesco et al., 1982), AMICA complemented the spectrum of the shorter wavelength of NIRS. Knowing the distance to the LIDAR footprint improved the knowledge of the position of the Hayabusa spacecraft relative to Itokawa's mass center, using AMICA data, LIDAR measurements, and the Itokawa shape model (Gaskell et al., 2008; Abe et al., 2006b).

We attempted to observe α Sco with both AMICA and NIRS in the IOP. During the long (134-s) AMICA exposure, a portion of sky in the vicinity of α Sco was scanned with NIRS to check its geometric calibration. The boresight of NIRS was centered on pixel (493, 510) in the AMICA frame. The alignment between NIRS and LIDAR was adjusted as accurately as possible on the ground prior to the launch. Because LIDAR (a 0.04°x0.97° ellipse) spectra reflected by Itokawa were detected by NIRS (0.1°x 0.1° squared), the alignment accuracy was better than the field-of-view of NIRS, i.e., ~0.1° (Abe et al. 2006b).

### 4-1-2. Focal length and distortion

Wide-band images of a portion of sky in Taurus were analyzed to determine the focal length $f$ (mm) and distortion (parameterized by $K_1$). A long-exposure image (178 s) was taken on 26 April 2005. Due to limited observation time, we could not acquire long-exposure images using the other seven filters. In this image, we identified 69 stars listed in the Hipparcos and Tycho catalogs, with negligibly small position errors. The position of each star in AMICA coordinates $(H^{(i)}, V^{(i)})$ was obtained by fitting a two-dimensional Gaussian function. The position accuracy, which depends on the ratio of the star signal to the background noise, was equivalent to 0.1–0.3 pixels.

The position of the $i$-th star in the catalogs $(\alpha^{(i)}, \delta^{(i)})$ can be converted into AMICA CCD coordinates $(X^{(i)}_{star}, Y^{(i)}_{star})$ in mm as follows:

$$\begin{pmatrix} X^{(i)}_{star} \\ Y^{(i)}_{star} \end{pmatrix} = \mathrm{C}\frac{L_{PIX}}{\Delta_{\mathrm{PIX}}}\begin{pmatrix} \cos\theta & -sin\,\theta \\ sin\,\theta & \cos\theta \end{pmatrix}\begin{pmatrix} -\dfrac{\cos\delta^{(i)}\, sin(\alpha^{(i)}-\alpha_0)}{\cos\delta_0\cos\delta^{(i)}\cos(\alpha^{(i)}-\alpha_0)+sin\,\delta_0\, sin\,\delta^{(i)}} \\ -\dfrac{sin\,\delta_0\cos\delta^{(i)}\cos(\alpha^{(i)}-\alpha_0)-\cos\delta_0\, sin\,\delta^{(i)}}{\cos\delta_0\cos\delta^{(i)}\cos(\alpha^{(i)}-\alpha_0)+sin\,\delta_0\, sin\,\delta^{(i)}} \end{pmatrix}, \qquad (2)$$

where C = 648000/π arcsec; $\alpha_0$ and $\delta_0$ are the right ascension and declination of the center of the AMICA image ($H$=512, $V$=512) in zero-based coordinates; and $\theta$ is the angle of rotation around the optical axis of AMICA using the right-hand rule. We assumed the optical axis of AMICA was co-incident with the center of the AMICA image and perpendicular to the CCD plane. $\Delta_{PIX}$ and $L_{PIX}$ (=1.2x10$^{-2}$ mm pixel$^{-1}$) are the pixel angular resolution in arcsec pixel$^{-1}$ and the size of a pixel in mm pixel$^{-1}$, respectively. The distortion parameter $K_1$ follows the scheme of Davies et al. (1994) and Klaasen et al. (1997), as follows:

$$\begin{cases} x^{(i)} = L_{PIX}(H^{(i)} - 512) \\ y^{(i)} = L_{PIX}(V^{(i)} - 512) \end{cases}$$

$$R^{(i)2} = x^{(i)2} + y^{(i)2} \tag{3}$$

$$\begin{cases} x_C^{(i)} = x^{(i)}(1 + K_1 R^{(i)2}) \\ y_C^{(i)} = y^{(i)}(1 + K_1 R^{(i)2}) \end{cases},$$

where $x^{(i)}_C$ and $y^{(i)}_C$ are the observed star offsets from the image center, corrected for radial distortion. The observed stellar positions ($x^{(i)}_C$, $y^{(i)}_C$) in the focal plane were compared with those of catalog stars ($X^{(i)}_{star}$, $Y^{(i)}_{star}$) in Eq. (2). A least-squares algorithm was solved for pixel size, camera orientation, and the distortion coefficient. The results are summarized in Table 4. The best-fit focal length was 120.71 ± 0.03 mm and the best-fit distortion coefficient $K_1$ was $-2.8\times10^{-5}$ mm$^{-2}$. The mean residual for the best solution was $3.6\times10^{-3}$ mm, or 0.3 pixels. When we ignored the distortion term (i.e. $K_1=0$ in Eq. (3)), the resulting pixel scale was 20.490 ± 0.005 arcsec and the effective focal length was 120.80 ± 0.03 mm. The resultant focal length was slightly longer than the designed value of 120 mm, but consistent with results obtained from pre-launch laboratory data (~120.6 mm).

[Table 4]

Note that the focal length and distortion were examined in only wide-band filter. The main engine of the spacecraft, ion engine had always disturbed the attitude of the spacecraft, which made us difficult to take images during MP. As the result, we could not examine the distortion and the focal length in the other seven filters.

**4-1-3. Alignment to the spacecraft**

Here we examine the alignment of AMICA with respect to the spacecraft. We used the results of star identification in nine images from the 2005 August and 2005 September runs to derive the relative alignment of the boresight of AMICA to the spacecraft. The rotation angles were estimated to adjust the observed position of stars to those of catalog stars. A set of rotation angles about the X- and Y-axes and a rotation angle about the Z-axis, which is identical to $\theta$ in Eq. (2), were adjusted separately but iteratively. The root-mean square (RMS) residual errors of the star positions in a single image were less than 0.51 pixels or 0.003°. The best estimates of the rotation angles from the AMICA coordinate to the spacecraft coordinate were -0.124° ± 0.011°, -0.087° ± 0.005°, and 0.380° ± 0.015° about the X-, Y-, and Z-axes, respectively. The estimated alignment includes errors from the spacecraft attitude determination, expected to be below 0.02°. Because the discrepancies among individual results from each star image are comparable to the uncertainty in spacecraft attitude, we conclude that most of the alignment errors arise from the uncertainty in spacecraft attitude measurements. All of the geometric information will be made available as a NAIF/SPICE (spacecraft position, planet/satellite position, instrument alignment, camera/scan platform instrument and spacecraft pointing, and events) instrumental kernel (Acton, 1996) in both the ISAS and PDS archives.

### 4-2. Radiometric calibrations

In this section, we describe the processes used to remove the instrumental effects and convert the observed raw Digital Numbers (DNs) into radiance units using mainly in-flight test data. The number of photons detected by AMICA CCD is converted to the number of electrons by the photon gain and the electrons are converted to DN by the electronic gain (17 $e^-$/DN for AMICA). Figure 5 provides a schematic illustration of the AMICA calibration pipeline. The pipeline converts raw DN unit into the physical units. As described below, we deduced the conversion factor in radiance units for v-band and the conversion factor in reflectance for ul-, b-, v-, w-, x-, p-, and zs-bands based on the inflight data (see Note and Limitation in Section 1. Introduction)

[Figure 5]

### 4-2-1. Bias and Dark Subtraction

The optical sky background signal level is a combination of an electronic offset that defines the “zero” level of the analog-to-digital converter (bias) and any additional signal due to thermal electrons (i.e., dark current). Because AMICA comprises two different areas (the imaging section and the store section), the dark current is the sum of these two components. A pre-flight dark current model was developed by fitting empirical forms using data from dark frames taken at multiple temperatures (-30 ~ +40°C). The dark current accumulation rate $I_{\mathrm{d,IM}}$ $(H,V,T_{\mathrm{CCD}})$ [DN $s^{-1}$] for each pixel can be expressed as follows:

$$I_{\mathrm{d,\,IM}}(H,V,T_{\mathrm{CCD}}) = d_{0,IM}(H,V)\exp\left[d_{1,IM}(H,V)T_{\mathrm{CCD}}\right], \qquad -30 \le T_{\mathrm{CCD}} \le +40 \quad , \tag{4}$$

where $d_{0,\,\mathrm{IM}}$ and $d_{1,\,\mathrm{IM}}$ are dark model parameters in the image section. Figure 6(a) shows the CCD temperature dependence of average dark current of all unmasked pixels (1000 columns x 1024 rows). To derive the dark current model in the store section, we use 1-μs exposure dark images. Because the readout time depends on CCD position (16 ms at (H,V)=(0,0) and 391 ms at (1024,1024)), the dark current in the store section generated a smooth brightness gradient across the 1-μs exposure frames. By fitting this gradient, we obtained the average value of dark accumulation rate in the store section as follows:

$$I_{\mathrm{d,\,ST}}(T_{\mathrm{CCD}}) = d_{0,ST}\exp\left(d_{1,ST}T_{\mathrm{CCD}}\right), \qquad -30 \le T_{\mathrm{CCD}} \le +40 \tag{5}$$

where $d_{0,\mathrm{ST}}$ and $d_{1,\mathrm{ST}}$ are dark model parameters in the store section. In Table 5, we summarize the pre-flight dark model parameters for average $I_{\mathrm{d,IM}}(T_{\mathrm{CCD}})$ and $I_{\mathrm{d,ST}}(T_{\mathrm{CCD}})$.

[Figure 6a]

[Figure 6b]

[Table 5]

After launch, we could not update our dark model using in-flight data because the CCD temperature was too low to detect dark current. Figure 7 shows the history of $T_{\mathrm{CCD}}$ after the launch. It was kept between -18°C (the descending and ascending period) and –30°C. Using Eq. (4) and (5), the dark accumulation rate averaged over the frame was estimated to be 0.16 and 0.15 [DN $s^{-1}$] at $T_{\mathrm{CCD}}$ = -18°C and $T_{\mathrm{CCD}}$ = -30°C, respectively. Therefore, any in-flight dark current was buried in the readout noise (60$e^-$ or ~4DN for AMICA). In the Itokawa observations, the exposure time was <<1 s, while Itokawa's intensity was 1000–3000 DN. Therefore, we conclude that dark current was negligible in the Itokawa images.

[Figure 7]

During flight, the CCD was exposed to severe radiation, and the damage from cosmic-ray irradiation resulted in an increase in the number of pixels with a larger than normal dark current, or "hot" pixels (Sirianni et al., 2005). Figure 8 shows a histogram of dark sky taken on 22 May 2003, 13 January 2004 and 22 August 2005. The histogram shows that the number of hot pixels increased during the mission. These hot pixels make accurate measurements of intensities of faint objects (e.g., stars) difficult. Hot pixels were also observed in the Itokawa images using less-sensitive filters (ul and zs-bands). We identified hot pixels using the optical navigation images taken during the AP and subtracting them from the Itokawa images in which the dark signal was large (i.e., $I_{\mathrm{DARK}}(\mathrm{H,V}) \times t_{\mathrm{EXP}} > 30$ DN). For reference, Table 6 shows the positions and dark currents of the top five hot pixels.

[Figure 8]

[Table 6]

Here we investigated the background zero level (i.e. bias). The bias level generally depends on the electronics temperature (Bell et al. 2003, 2006) and CCD temperature. From pre-flight standalone experiments, we found empirically that the bias level increased when the analog electronics (ONC-AE) temperature decreased. However, we could not establish the bias model as a function of temperature because ONC-AE was not equipped with a temperature sensor. Figure 9 shows the bias level history, obtained from blank sky data. It is clear that the bias level showed secular variation, and thus we fit the bias level as a function of time, as follows:

$$I_{BIAS}(DAY) = B_0 + B_1 DAY + B_2 DAY^2 \quad , \tag{6}$$

where $DAY$ is the number of days after the launch, and $B_0$, $B_1$, and $B_2$ are in-flight bias model parameters. The best-fit parameters are given in Table 7. Reproducible error in the bias level was ≤ 5 DN (1σ). We emphasize that the bias uncertainty can be corrected eventually in most of the Itokawa frames by adjusting the sky level to zero DN, because almost all Itokawa images (except close-up images) include the dark sky area.

[Figure 9]

[Table 7]

**4-2-2. Response Uniformity and Temporal Variation**

Flat-field images in all bands were acquired on the ground to correct pixel-to-pixel variation in CCD response and vignetting (reduction of image brightness at the periphery compared to the center). Standalone flat-field data were obtained twice in each filter using an integrating sphere at NEC Toshiba Space Systems Ltd. (NT Space) at room temperature (around 30°C). AMICA was pointed into the integrating sphere to acquire images of a field known to be spatially uniform to an accuracy of ~2%. A low spatial frequency responsivity variation with amplitudes of 4% was found, as well as pixel-to-pixel responsivity variation with amplitudes of 0.5% or less and 1–10 pixel-scale dark spots. The low spatial frequency component can be interpreted as vignetting: a decrease in detected image intensity with increasing distance from the image center. The highest spatial frequency (pixel-to-pixel) components represent nonuniformity in pixel sensitivity. Dark spots with several-pixel scale are attributed to dust on the light path or defective pixels. In addition, we found a doughnut-shaped pattern in the NT Space integrating sphere of amplitude 3% or less (Fig. 10). We considered that the doughnut-shaped pattern was caused by the dark spot in the integrating sphere. Because the uneven pattern appeared at different CCD positions in each test, we masked the pattern and added two images, excluding the masked region. Furthermore, we developed a small integrating sphere, which could be used in the synthesis test (Sogame et al. 2005). The spatial uniformity of the small integrating sphere matched the former integrating sphere to an accuracy of 1%, except the doughnut-shaped pattern. We obtained flat images of the portable integrating sphere in w- and p-band filters during a functional test of the spacecraft. The archived flat-field images are derived from the NT Space integrating sphere data for ul-, b-, v-, x-, zs-, and wide-band filters, and from the portable integrating sphere for w- and p-band filters.

[Figure 10]

Post-launch, no photometrically flat objects were available in space to use in assessing ongoing variation in the flat-field calibration. To check the temporal variation in responsivity, two small lamps (commonly called FF-lamp A and B) were attached in front of the objective on AMICA (see Fig. 2 and Fig. 4). By activating these FF-lamps in high- or low-voltage mode, we obtained images of defocused light to check the camera conditions. Figures 11(a) and (b) are w-band images taken by the integrating sphere and FF-lamps. All of the dark spots in the flat-field image in Figure 11(a) appeared in the FF-lamps image shown in Figure 11(b), although differences in shapes and absorption rates are apparent. In the FF-lamps images, we found twinned dark spots aligned on the lines parallel to the FF-lamp A and B line connection; these spots are not visible in the pre-flight flat-field images. Although their cause has not been extensively investigated, they are probably of FF-lamp origin.

Figures 11(c) and (d) are FF-lamp images taken during in-flight operations. We show Figures 11(c)–(e) to trace the time variation in the FF-lamp image data. Figure 11(e) is the ratio of (c) to (b), and Fig. 11(f) is the ration of (d) to (c). Each contour level corresponds to a 1% variation. In Figure 11(e), intensities of some pixels varied by up to ±3% due to temporal variation in the twinned spots. The change can be explained by the attachment and removal of dust particles or the evaporation of contaminants during the launch phase. The variation in FF-lamp images do not necessarily indicate that the flat-fielding varied by up to 3% because the twinned spots were not obvious in the flat-field image. Many circular spots were generated in the CP (see Fig. 11(f)), but the variation in intensity was ≤ 1%. FF-lamp images were obtained during the MP. However, one of the FF-lamps did not turn on because of a breakdown. Further comparative studies would be needed to convert the FF-lamp images into a flat-field image if one consider the pixel-to-pixel responsivity less than ~3%. AMICA data users should pay particular attention to pixel values around (380, 450), (390, 510), (730, 390), (830, 920), and (850, 510), where temporal variation in the FF-lamp images are larger than 2% (Fig. 11(e)).

[Figure 11]

The lamps had high- and low-voltage modes; we used the low-voltage mode for the longer wavelength bands (x, p, zs) to avoid saturation. Figure 12 shows the temporal variation in the raw AMICA output (average in the H = 462-561 and V = 462-561 square) during both the CP and MP. The x-, p-, and zs-band data output (dashed lines in Fig. 12) exhibited large variation, possibly attributable to the instability of lamp brightness in the low-voltage mode. In contrast, the ul-, b-, v-, and w-band data showed little variation (solid lines in Fig. 12). Applying the smear correction, the mean intensity of the v- and w-band showed no change to an accuracy of 1%, but the mean values of the b-band decreased slightly (~3%). This tendency met manufacturer expectancy that degradation was insignificant at $\lambda$>390 nm. In addition, the result of the degradation estimated by the FF-lamp is consistent with that calculated from observations of $\tau$ Sco (the responsivity in b-band was estimated to be down by ~5% by stellar photometry). Therefore, we conclude that the calibration factors in v-band determined by stellar observations during CP are applicable to Itokawa data. We deduced the conversion factors in the other six channels (ul-, b-, v-, w-, x-, p-, and zs-bands) by the observations of Itokawa itself to avoid the effect of degradation. Although FF-lamps were not originally designed for radiometric calibration, we eventually found that they were useful in checking the degradation.

[Figure 12]

**4-2-3. Readout Smear**

As we mentioned in Section 2, AMICA is shuttered electronically. Observed images are exposed for the designated exposure time and the vertical charge-transfer period ($t_{VCT}$ = 12μs×1024 = 12.288 ms). The finite duration of frame transfer produces a streak (readout smear) parallel to the direction of frame transfer (along the V-direction). The middle image in Figure 13b shows an example image before smear subtraction. The readout smear is obvious in images with exposure times shorter than ~100 μs.

In a nominal imaging sequence, we planned to take a pair of images over a short time interval (typically 15 s) and subtract the latter frame (smear frame, 0 s exposure) from the former frame (≥ 5.44-ms exposure) on the ONC-E. After an anomaly in the $Y_{SC}$–reaction wheel on 2 October 2005, the spacecraft attitude could not be adequately controlled, which caused a mismatch in pointing between Itokawa and smear image pairs (Fig. 13(a)). Because of the pointing mismatch, this image pair subtraction technique could no longer be used to remove the streaks on the observed images. As a result, the following smear model was developed and used to remove smear from images taken during the descending and ascending period.

The readout smear brightness $I_{SMEAR}(H)$ can be modeled from the observed images in a way similar to Murchie et al. (1999) and Bell et al. (2006), as follows:

$$I_{\rm SMEAR}(H) = \sum_{H=0}^{N_{\rm V}-1} \frac{t_{\rm VCT}}{t_{\rm VCT} + t_{\rm EXP}} \frac{I_{\rm RAW}(H,V) - I_{\rm SKY}(H,V)}{N_{\rm V}} \quad , \tag{7}$$

where $I_{\rm RAW}(H, V)$ is the intensity of the raw data taken with the designated exposure time $t_{\rm EXP}$; $N_{\rm V}$=1024 is the pixel number along the V-direction; and $I_{\rm SKY}(H, V)$~300 DN is the sum of the bias and dark current (mainly in the store section of the CCD). In the right-hand images of Figure 13a, corrected by the smear model, the readout smear has been subtracted from the original image with a residual of less than 1 DN.

[Figure 13(a)]

[Figure 13(b)]

**4-2-4. Linearity**

DN values stored in raw images are approximately proportional to the number of photons detected in each CCD pixel during the exposure time. Linearity, the degree to which the CCD response is proportional to the incident flux, was tested during the pre-flight testing and the MP. The pre-flight linearity was examined by taking a spectralon plate illuminated by a halogen lamp at room temperature. We compared the DN values taken at different exposure times and found that intensity was linear with an incident flux between 1000 DN and 3500 DN (Fig. 14). By fitting the pre-flight data, we obtained the following:

$$I_{output} = I_{input}^{\gamma} + L_0 I_{input} \exp(L_1 I_{input}) \quad , \qquad (8)$$

where $I_{inpu}t$ and $I_{output}$ are the intensity of the spectralon and the observed intensity; $\gamma$=1-5.0x10$^{-8}$ is the power index of the output signal; $L_0$= -4.87x10$^{-11}$; and $L_1$= 5.09x10$^{-3}$ in the second term corrects the nonlinearity around the saturation limit ($I_{input}$> 3800 DN).

[Figure 14]

The in-flight linearity was investigated in the MP by imaging Itokawa through the v-band filter at exposure times of <1 μs (smear), 10.9 ms, 21.8 ms, 43.5 ms, 65.5 ms, 87.0 ms, 131.0 ms and 174.0 ms. For each image in the sequence, the observed flux was measured by averaging over 20-pixel by 20-pixel boxes centered on brighter terrain and darker terrain. The results are shown in Figure 15(a). The obtained flux was linear with exposure times less than 131.0 ms in the darker terrain, but was not linear beyond 131.0 ms in the brighter terrain. The obtained data were corrected for readout smear and background level, and converted into accumulation rates per unit of time (DN $ms^{-1}$) for more precise investigations. The dark accumulation rate was constant to an accuracy of 0.3% when the signal level ranged between 780 DN and 3230 DN (Fig. 15(b), 10.9- to 131-ms exposures in the darker area), and dropped by 2% at a signal level of 3840 DN (131-ms exposure in the brighter area in Fig. 15(b)). This result was consistent with pre-flight investigations. Therefore, we applied Eq. (8) to the Itokawa linearity correction.

[Figure 15(a)]

[Figure 15(b)]

**4-2-5. Radiometric Responsivity**

Due to the finite width of the AMICA filters, calibration factors were affected by instrumental system efficiency (Fig. 3) and by the spectrum of the incident light sources. The best approach for determining the radiometric calibration coefficients is to use objects whose colors are similar to Itokawa. Stars listed in ECAS standard star catalog meet this criterion not only because their magnitudes and color indices are well known, but also because their spectra are similar to those of solar system objects. However, the installation of the ND filter made it difficult to use ECAS standard stars with magnitudes fainter than 5.96. $\alpha$ Crv (F0), which was observed by AMICA in the MP, is a good AMICA calibration star candidate. Because the magnitude and color indices of $\alpha$ Crv are unknown, we are planning to observe it through ECAS filters from a ground-based observatory. In this paper, we determined the calibration coefficient from DN to the radiance for v-band stellar magnitudes.

[Table 8]

We observed nine stars in the AMICA v-band filter. In Table 8, we summarize the spectral type, observed date, and Johnson system's V-band magnitudes. In addition, we list the maximum and minimum magnitude if the stars appeared in the variable star catalogs (General catalogue of Variable Stars, GCVS, Samus' et al., 2002 and catalogue of suspected variable stars,

Kazarovets et al., 1998). Because the magnitude of α Ori changes significantly, we removed it from the reference stars list. We include variable stars other than α Ori whose amplitudes seem to be less than ± 0.05 mag. (5%). Figure 16 shows the relationship between Johnson system's catalog magnitude $V_{cat}$ (error bars denote variable star amplitudes) and the AMICA instrumental magnitude $v_{inst}$ defined as follows:

$$v_{inst} = -2.5\log\left(\frac{I_{star}}{t_{exp}}\right) , \qquad (9)$$

where $I_{star}$ is the observed v-band intensity (DN $s^{-1}$). The v-band instrumental magnitude showed good correlation with the V-band catalog magnitude and the deviations from the fit-line are independent of the spectra of the observed nine stars. This result was expected, as the transmittance of the AMICA v-band filter was designed to match that of the Johnson V-band filter. From the linear fit, we obtained the magnitude zero point 7.55 ± 0.03. To convert the magnitude into flux, the absolute flux of Vega (Bohlin and Gilliland, 2004) was weighted by the AMICA system efficiency (Fig. 3(b)), and the effective flux of a zero-magnitude star was calculated as $3.53\text{x}10^{-8}$ W $m^{-2}$ $\mu m^{-1}$ in AMICA v-band. This calculated flux is comparable to $3.61\text{x}10^{-8}$ W $m^{-2}$ $\mu m^{-1}$ (Bessel 1979) in the Johnson V-band to the accuracy of ~2%. Applying the calculated values, we obtained an AMICA v-band calibration factor of $(3.42 \pm 0.10)\times10^{-3}$ [(W $m^{-2}$ $\mu m^{-1}$ $sr^{-1}$) / (DN $s^{-1}$)].

[Figure 16]

The color indices of Itokawa at a phase angle (Sun-Itokawa-AMICA angle) of 21° were obtained using the University of Hawaii 2.2-m telescope (UH88) by David J. Tholen to an accuracy of <1%. He could not obtain the data in z-band due to the faintness. Using his observational data, it was possible to determine the conversion factor for filters other than v-band, to match the AMICA measurements to the ground-based observations. However, it is important to note that the color indices are phase angle-dependent. We used the disk-integrated Itokawa data on 11 October 2005, at a solar phase angle of ~21°. By comparing the disk-integrated Itokawa DN with ground-based data, we deduced the conversion factor from DN to reflectance relative to the v-band (Table 9). Figure 17 shows the comparison between spectra taken at different rotational phase (a) and different times of day (b). Although there are small time fluctuations in the AMICA spectra up to ~1%, they are eventually in excellent agreement with those obtained by the ground-based observatories, i.e., the Palomar Observatory (<0.55μm, Lowry et al. 200) and the Kitt Peak and IRTF observatories (Binzel et al. 2001).

[Table 9]

[Figure 17]

**4-3. Scattered Light and Point-Spread Functions (PSFs)**

The point-spread function (PSF) describes the response of an imaging system to a point source (see e.g. Klaasen et al. 1997). It may be affected by optical system aberrations (including lenses, filters and the CCD surface), and therefore the PSF is likely to be wavelength-dependent. We examined the pixel-scale PSF (less than 10 pixels) using bright point sources (α Sco, β Sco, τ Sco, α Aur, α Ori, β Tau in the CP; Itokawa in the AP; and α Vir in the MP) in various positions on the CCD chip (0°–2.3° away from the image center). Figure 18 shows the PSFs obtained in the AP as well as the CP. The object centers are computed by fitting a Gaussian to the marginal profiles in x and y using non-linear least-squares techniques. The PSFs were independent of CCD chip position and observation time, but showed a weak dependency on filter type.

[Figure 18]

In addition to the pixel-scale PSF, a low-intensity light component was observed in images of the Moon and Itokawa in which the PSF extent was greater than 10 pixels. During the Earth swing-by in May 2004, we found a dim halo around the Moon. This feature was observed most clearly next to bright light sources, and as discussed in previous studies (e.g., Gaddis et al., 1995; Murchie et al., 2002), may have originated from light scattered inside the camera and stray light outside the camera FOV. The intensities of halos were wavelength-dependent, with brighter halos observed in zs-band and darker halos in b-, v-, and w-bands (Fig. 19). Because no pre-flight data of scattered light were acquired to correct the scattered light and stray light, we plan to examine the PSFs using the Moon and Itokawa images and correct for them using the image deconvolution method (see Gaddis et al., 1995).

[Figure 19]

In this paper, we focus on improvements in resolution using the pixel-scale PSFs shown in Figure 18. All observed images were blurred due to the PSFs. In general, image-restoration techniques provide modest gains in resolution, remove scattered light and stray light, and improve the signal-to-noise ratio. Several image deconvolution algorithms are available, such as the Richardson–Lucy algorithm, the Wiener filter, and the s-CLERAN algorithm. We applied the Richardson–Lucy algorithm (hereafter, RL algorithm; Richardson 1972; Lucy 1974, see also White 1994), which converges to the maximum likelihood solution for Poisson statistics in the data. The noiseless blurred image can be written as follows:

$$I_k(i) = \sum_j P_k(i \mid j) O_k(j) \quad , \tag{10}$$

where $P_k(i \mid j)$ denotes the PSF at a given filter $k$, that is, the fraction of light from the true location $j$ that is scattered into observed pixel $i$; and $O$ is the unblurred object. The RL algorithm takes the image estimates and improves them using the following iteration equation:

$$O_{k,n+1}(j) = O_{k,n}(j) \frac{\sum_i P_k(i \mid j) \frac{D_k(i) + R^2/g}{I_{k,n}(i) + R^2/g}}{\sum_i P_k(i \mid j)} \quad , \tag{11}$$

where $D(i)$ is the observed image; $I_n$ is the $n$-th estimate of the blurred image; and $R$ and $g$ are the read-out noise and gain (number of electrons per DN), respectively. After preparing the images by correcting hot pixels using the in-flight hot pixel table and the axisymmetric PSFs shown in Fig. 18 plus the read-out noise model (Table 1), iteration of Eq. (8) generated the restored images. Figure 20 shows an example resultant image.

[Figure 20]

The image restoration algorithm yielded good results, improving the image resolution in all AMICA channels.

## 4-4. Stray light

The dark level was monitored in-flight during star observations; that is, the backgrounds of the star images were regarded as the dark level, composed of dark current and bias. The brightness of the natural sky background in the AMICA viewing direction ($2 \times 10^{-6}$ W $m^{-2}$ str μm at 0.55 μm, Leinert et al., 1998) is too faint to be detected with AMICA (estimated intensity ~$1 \times 10^{-3}$ DN $s^{-1}$). Nevertheless, the background sky level of several observations was higher than expected, although the CCD temperature stayed within a small temperature range (-29 to -26ºC). These higher levels were probably due to light from the off-optical axis, incompletely blocked by the AMICA baffle.

In the nominal attitude of the spacecraft, the Sun shines on $+Z_{SC}$ and $+X_{SC}$ panels, while AMICA is attached on the $-Z_{SC}$ panel near the $-X_{SC}$ plane (Fig. 21). We found that data with higher dark levels were obtained when the Sun illuminated $\pm Y_{SC}$ panels. In this tilted position, the Sun could illuminate the sampler horn, which is located close to AMICA. We examined the light rate [DN $s^{-1}$] as a function of an azimuth angle $\phi$, the angle between the $+X_{SC}$ direction of the spacecraft and the Sun-pointing vector projected on the spacecraft $X_{SC}$–$Y_{SC}$ plane (see Fig. 21). The tilt angle of the $Z_{SC}$–$Y_{SC}$ plane, $\theta$, was limited to within 10˚. Figure 22 supports the conjecture that the light rate varies with varying $\phi$. This limitation of $\theta$ ensured that the stray light contribution would be less than ~0.1 DN for Itokawa observations because the asteroid observation exposures would be approximately 0.1 s or less. Therefore, we ignored off-axis stray light from the Sun in Itokawa images.

[Figure 21]

[Figure 22]

## 5. Summary

In this paper, we outlined the design, in-flight operation, performance, and calibration of AMICA. Key in-flight calibration results include:

(1) The AMICA signal is linear with respect to input signal to an accuracy of << 1% when the signal level is < 3800 DN.

(2) The AMICA v-band (0.55 μm) absolute radiance calibration is accurate to within 4% or less.

(3) The pixel-to-pixel precision of the calibrated relative reflectance data is less

than 3%.

(4) The accuracy of the disk-integrated spectra with respect to the AMICA v-band is 1% or less.

(5) The background zero level uncertainty is ≤ 5 DN.

(6) Both dark current and stray light from outside the FOV are negligibly small in the Itokawa images.

(7) The focal length is 120.80 ± 0.03 mm (distortion parameter ignored), yielding a 5.83°×5.69° FOV.

The calibration provided a geometric albedo for Itokawa of 0.24 ± 0.02 (Ishiguro et al, in preparation), in good agreement with that obtained from ground-based observations (Thomas-Osip et al., 2008). The calibrated AMICA spectra are also well-matched the spectra taken by the ground-based observatories. Scattered light inside the optics, most obvious in the p and zs-bands, was identified, but not yet quantified to permit removal from image data. We plan to improve the point-spread function to subtract the components by image deconvolution.

**Appendix. ONC File Naming Conventions**

The file naming convention for ONC images is {*T*}{*C*}{*nnnnnnnnnn*}_{*filter*}.fits, where: {*T*} indicates the type of data ("S" for scientific data or "N" for navigation data), {*C*} identifies the name of the instrument (T, W1, and W2 for ONC-T (i.e. AMICA), ONC-W1, and ONC-W2, respectively); {*nnnnnnnnnn*} is the mission elapsed time when the images is generated in ONC-E; {*filter*} is the name of the filter (ul, b, v, w, x, p, and zs, respectively).

**Acknowledgments**

The detailed inflight testing of the AMICA was made possible by the dedication and hard work of many members of the Hayabusa Mission teams. We especially express deepest appreciation to M. Nagano and the former AMICA PI Dr. T. Nakamura, not only for their strenuous effort on the development of ONC/AMICA but also for their invaluable advices and encouragement. We also thank to the NEC Toshiba Space and NEC Aerospace Systems engineers, K. Noguchi, H. Hihara, H. Hosoda and S. Kanda for their efforts in developing the ONC/AMICA, and T. Kominato, M. Matsuoka, K. Shirakawa, M Uo and O. Oshima for supporting the inflight operations. Some of Hayabusa team members, especially Drs. F. Yoshida, B. Dermawan, and S. Hasegawa helped during preflight testing. Discussions with Professors M. Abe, T. Okada M. Yoshikawa, A. Fujiwara, and J. Kawaguchi were very valuable. We thank two reviewers, Brian Carcich and Mark Robinson, for their critical reading of the manuscript and their helpful comments. Multi-color data in the ECAS photometric system was obtained under the support with Institute for Astronomy, University of Hawaii. The staff in Bosscha observatory, Institute Technology of Bandung also supported our ground-based observation for calibration of AMICA. M. Ishiguro was supported by JSPS Postdoctoral Fellowships for Research Abroad, National Astronomical Observatory of Japan, and Seoul National University.

## References

Abe, M., Y. Takagi, K. Kitazato, T. Hiroi, S. Abe, F. Vilas, B. E. Clark, and A. Fujiwara, 2005. Observations with near infrared spectrometer for Hayabusa mission in the Cruise phase. Lunar Planet. Sci. 36, Abstract No. 1604

Abe, M., Y. Takagi, K. Kitazato, S. Abe, T. Hiroi, F. Vilas, B. E. Clark, P. A. Abell, S. Lederer, K. S. Jarvis, T. Nimura, Y. Ueda, A. Fujiwara 2006a. Near-infrared Spectral Results of Asteroid Itokawa from the Hayabusa Spacecraft. *Science* **312**, 1334-1338.

Abe, M., Vilas, F., Kitazato, K., Abell, P. A., Takagi, Y., Abe, S., Hiroi, T., Clark, B. E. 2007. I n-Flight Calibration of the Hayabusa Near Infrared Spectrometer (NIRS). Lunar Planet. Sci. 38, Abstract No. 1338.

Abe, S., T. Mukai, N. Hirata, O. Barnouin-Jha, A. F. Cheng, H. Demura, R. W. Gaskell, T. Hshimoto, K. Hiraoka, T. Honda, T. Kubota, M. Matsuoka, T. Mizuno, R. Nakamura, D. J. Scheeres, M. Yoshikawa 2006b. Mass and Local Topography Measurements of Itokawa by

Hayabusa. *Science* **312**, 1344-1347.

Acton, C. H., Ancillary data services of NASA's Navigation and Ancillary Information Facility, *Planetary and Space Science* **44**, 65–70, 1996.

Bell, J. F., Joseph, J., Sohl-Dickstein, J. N., Arneson, H. M., Johnson, M. J., Lemmon, M. T., Savransky, D. 2006. In-flight calibration and performance of the Mars Exploration Rover Panoramic Camera (Pancam) instruments. *Journal of Geophysical Research*, Volume 111, Issue E2, CiteID E02S03.

Bessell, M. S. 1979. UBVRI photometry. II - The Cousins VRI system, its temperature and absolute flux calibration, and relevance for two-dimensional photometry. *Publications of Astronomical Society of the Pacific*, **91**, 589-607

Binzel, R. P., Rivkin, A. S., Bus, S. J., Sunshine, J. M., Burbine, T. H. 2001. MUSES-C target asteroid (25143) 1998 SF36: A reddened ordinary chondrite. *Meteoritics and Planetary Science* **36**, 1167-1172.

Bohlin, R. C., Gilliland, R. L. 2004. Hubble Space Telescope Absolute Spectrophotometry of Vega from the Far-Ultraviolet to the Infrared. *Astronomical Journal*, **127**, 3508-3515.

Davies, M. E., Colvin, T. R., Belton, M. J. S., Veverka, J., Thomas, P. C. 1994. The direction of the north pole and the control network of Asteroid 951 Gaspra. *Icarus*, **107**, 18-22.

Demura, H., S. Kobayashi, E. Nemoto, N. Matsumoto, M. Furuya, A. Yukishita, N. Muranaka, H. Morita, K. Shirakawa, M. Maruya, H. Ohyama, M. Uo, T. Kubota, T. Hashimoto, J. Kawaguchi, A. Fujiwara, J. Saito, S. Sasaki, H. Miyamoto, N. Hirata 2006. *Science* **312**, 5778, 1347-1349.

Fujiwara, A., J. Kawaguchi, D. K. Yeomans, M. Abe, T. Mukai, T. Okada, J. Saito, H. Yano, M. Yoshikawa, D. J. Scheeres, O. Barnouin-Jha, A. F. Cheng, H. Demura, R. W. Gaskell, N. Hirata, H. Ikeda, T. Kominato, H. Miyamoto, A. M. Nakamura, R. Nakamura, S. Sasaki, K. Uesugi 2006. The Rubble-Pile Asteroid Itokawa as Observed by Hayabusa. *Science* **312**, 1330-1334.

Fuse, T., Yoshida, F., Tholen, D., Ishiguro, M., Saito, J. 2008. Searching Satellites of Asteroid Itokawa by Imaging Observation with Hayabusa Spacecraft. *Earth, Planets and Space* **60**, 1, 33.

Gaddis, L. R., McEwen, A. S., Becker, T. L., 1995, Compositional variations on the Moon: Recalibration of Galileo solid-state imaging data for the Orientale region and farside. Journal of Geophysical Research, Volume 100, Issue E12, p. 26345-26356.

Gaskell, R. W., Barnouin-Jha, O. S., Scheeres, D. J., Konopliv, A. S., Mukai, T., Abe, S., Saito, J., Ishiguro, M., Kubota, T., Hashimoto, T., Kawaguchi, J., Yoshikawa, M., Shirakawa, K., Kominato, T., Hirata, N., Demura, H., “Characterizing and navigating small bodies with imaging data”, 2008, Meteoritics & Planetary Science **43**, 6, 1049-1061.

Hashimoto, T., T. Kubota, and T. Mizuno, Lightweight sensors for autonomous asteroid landing of Muses-C mission, 2003. *Acta Astronautica* **52**, 381–388

Hillier, J. K., B. J. Buratti, and K. Hill 1999, Multispectral photometry of the Moon and absolute calibration of the Clementine UV/Vis camera, *Icarus* **141**, 205–225.

Hirata, N., O. S. Barnouin-Jha, C. Honda, R. Nakamura, H. Miyamoto, S. Sasaki, H. Demura, A. M. Nakamura, T. Michikami, R. W. Gaskell, J. Saito 2009. A survey of possible impact

structures on 25143 Itokawa. *Icarus* **200**, 486-502

Hoffleit, D., Jaschek, C. 1991. The Bright star catalogue. New Haven, Conn.: Yale University Observatory, |c1991, 5th rev.ed., edited by Hoffleit, Dorrit; Jaschek, Carlos |v(coll.) .

Ishiguro, M., Hiroi, T., Tholen, D. J., Sasaki, S., Ueda, Y., Nimura, T., Abe, M., Clark, B. E., Yamamoto, A., Yoshida, F., Nakamura, R., Hirata, N., Miyamoto, H., Yokota, Y., Hashimoto, T., Kubota, T., Nakamura, A. M., Gaskell, R. W., and Saito, J. 2007. Global mapping of the degree of space weathering on asteroid 25143 Itokawa by Hayabusa/AMICA observations. *Meteoritics & Planetary Science* **42**, 10, 1703-1856.

Kawaguchi, J., H. Kuninaka, A. Fujiwara, and T. Uesugi, MUSES-C, its launch and early orbit operations, in *Proceedings of the Fifth IAA International Conference on Low-Cost Planetary Missions*, 2003. pp. 25–32, Noordwijk, The Netherlands, 24–26 September 2003

Kazarovets, E. V., Samus, N. N., Durlevich, O. V. 1998. New Catalogue of Suspected Variable Stars. Supplement - Version 1.0, Information Bulletin on Variable Stars, 4655, 1.

Kholopov, P. N.1996. General catalogue of variable stars. Vol.5. Moscow: Nauka Publishing House, |c1996, 4th ed., edited by Samus, Nikolai N.

Klaasen, K. P., Belton, M. J. S., Breneman, H. H., McEwen, A. S., Davies, M. E., Sullivan, R. J., Chapman, C. R., Neukum, G., Heffernan, C. M., Harch, A. P., Kaufman, J. M., Merline, W. J., Gaddis, L. R., Cunningham, W. F., Helfenstein, P., Colvin, T. R. 1997. Inflight performance characteristics, calibration, and utilization of the Galileo SSI camera. *Opt. Eng*., 36, 3001-3027

Leinert, Ch., Bowyer, S., Haikala, L. K., Hanner, M. S., Hauser, M. G., Levasseur-Regourd, A.-Ch., Mann, I., Mattila, K., Reach, W. T., Schlosser, W., Staude, H. J., Toller, G. N., Weiland, J. L., Weinberg, J. L., Witt, A. N. 1998. The 1997 reference of diffuse night sky brightness. *Astron. Astrophys. Suppl*. **127**, 1-99.

Li, H., Robinson, M. S., Murchie, S. 2002. Preliminary Remediation of Scattered Light in NEAR MSI Images. *Icarus*, **155**, 244-252.

Lowry, S. C., P. R. Weissman, M. D. Hicks, R. J. Whiteley, S. Larson 2005. Physical properties of Asteroid (25143) Itokawa—Target of the Hayabusa sample return mission. *Icarus* **176**, Issue 2,

408-417.

Lucy L. B. 1974. An iterative technique for the rectification of observed distributions. *Astronomical Journal* **79**, 745.

Michikami, T., A. M. Nakamura, N. Hirata 2009. The shape distribution of boulders on asteroid 25143 Itokawa: Comparison with fragments from impact experiments. *Icarus* (in press).

Mukai, T., H. Araki, T. Mizuno, N. Hatanaka, A. M. Nakamura, A. Kamei, H. Nakayama, and A. Cheng 2002. Detection of mass, shape and surface roughness of target asteroid of MUSES-C by LIDAR, *Advances in Space Research* **29**, 1231–1235.

Mukai, T., Nakamura, A. M., Sakai, T., 2006. Asteroidal surface studies by laboratory light scattering and LIDAR on HAYABUSA. *Advances in Space Research* **37**, 138-141.

Müller, T. G., T. Sekiguchi, M. Kaasalainen, M. Abe, S. Hasegawa 2005. Thermal infrared observations of the Hayabusa spacecraft target asteroid 25143 Itokawa. *Astronomy and Astrophysics* **443**, Issue 1, 347-355.

Murchie, S., M. Robinson, B. Clark, H. Li, P. Thomas, J. Joseph, B. Bussey, D. Domingue, J. Veverka, N. Izenberg, and C. Chapman 2002. Color variations on Eros from NEAR multispectral imaging, *Icarus* **155**, 145–168

Murchie, S., M. Robinson, S. E. Hawkins, A. Harch, P. Helfenstein, P. Thomas, K. Peacock, W. Owen, G. Heyler, P. Murphy, E. H. Darlington, A. Keeney, R. Gold, B. Clark, N. Izenberg, J. F. Bell, W. Merline, and J. Veverka 1999. Inflight calibration of the NEAR multispectral imager, *Icarus* **140**, 66–91.

Nakamura, R., Ishiguro, M., Nakamura, A. M., Hirata, N., Terazono, J., Yamamoto, A., Abe, M., Hashimoto, T., Saito, J. 2005. Inflight Calibration of Asteroid Multiband Imaging Camera Onboard Hayabusa: Preliminary Results. . Lunar Planet. Sci. 36, Abstract No. 1602.

Nakamura, T., A. M. Nakamura, J. Saito, S. Sasaki, R. Nakamura, H. Demura, H. Akiyama, and D. Tholen (AMICA Team) 2001. Multi-band imaging camera and its sciences for the Japanese near-Earth asteroid mission MUSES-C, *Earth, Planets and Space* **53**, 1047–1063.

Noguchi, T., A. Tsuchiyama, N. Hirata, H. Demura, R. Nakamura, H. Miyamoto, H. Yano, T. Nakamura, J. Saito, S. Sasaki, T. Hashimoto, T. Kubota, M. Ishiguro, M. E. Zolensky 2009. Surface morphological features of boulders on Asteroid 25143 Itokawa. *Icarus* (in press)

Richardson W. H. 1972. Bayesian-based iterative method of image restoration. *Journal of the Optical Society of America* **62**, 55-59.

Saito, J., H. Miyamoto, R. Nakamura, M. Ishiguro, T. Michikami, A. M. Nakamura, A. M., H. Demura, S. Sasaki, N. Hirata, C. Honda, A. Yamamoto, Y. Yokota, T. Fuse, F. Yoshida, D. J. Tholen, R. W. Gaskell, T. Hashimoto, T. Kubota, Y. Higuchi, T. Nakamura, P. Smith, K. Hiraoka, T. Honda, S. Kobayashi, M. Furuya, N. Matsumoto, E. Nemoto, A. Yukishita, K. Kitazato, B. Dermawan, A. Sogame, J. Terazono, C. Shinohara, H. Akiyama 2006. Detailed Images of Asteroid 25143 Itokawa from Hayabusa. *Science* **312**, 1341-1344.

Samus', N. N., Goranskii, V. P., Durlevich, O. V., Zharova, A. V., Kazarovets, E. V., Pastukhova, E. N.; Hazen, M. L., Tsvetkova, T. M. 2002. An Electronic Version of Volume I of the General Catalogue of Variable Stars with Improved Coordinates. *Astronomy Letters*, **28**, 174-181

Sogame, A., Saito, J., Hasegawa, S., Ishiguro, M. 2005. Report on the Development of the Small Integration Spheres for the Test of Asteroid Multiband Imaging Camera Mounted on HAYABUSA, and the Flat-field Test Using These Spheres at its Flight Operation Test (Japanese). *Journal of The Remote Sensing Society of Japan*, **25**, 372-378.

Sirianni, M., Mutchler, M., Lucas, R., 2005. Hot Pixels Growth in ACS CCDs. Proc. of 2005 HST Calibration Workshop, Space Telescope Science Institute, A. Koekemoer, Goudfrooij, P., and Dressel., L. eds., p. 1-6.

Tedesco E. F., D. J. Tholen, B. Zellner 1982. The eight-color asteroid survey - Standard stars. *Astronomical Journal* **87**, 1585-1592.

Thomas-Osip, J. E., Lederer, S. M., Osip, D. J., Vilas, F., Domingue, D., Jarvis, K., Leeds, S. L. 2008. The 2004 Las Campanas/Lowell Observatory Itokawa campaign: I. Simultaneous visible and near-infrared photometry of the Hayabusa mission target. *Earth, Planets, and Space* **60**, 39-48.

Yano, H., T. Kubota, H. Miyamoto, T. Okada, D. Scheeres, Y. Takagi, K. Yoshida, M. Abe, S.

Abe, O. Barnouin-Jha, A. Fujiwara, S. Hasegawa, T. Hashimoto, M. Ishiguro, M. Kato, J. Kawaguchi, T. Mukai, J. Saito, S. Sasaki, M. Yoshikawa 2006. Touchdown of the Hayabusa Spacecraft at the Muses Sea on Itokawa. *Science* **312**, 1350-1353

Yokota, Y., Ishiguro, M., Nakamura, A. M., Nakamura, R., Tholen, D., Smith, P., Saito, J., Kubota, T., Hashimoto, T. 2006. Opposition Effect on Itokawa: Preliminary Report from Hayabusa Images. Lunar Planet. Sci. 37, Abstract No. 2445.

White, R. L. 1994. Astronomical Data Analysis Software and Systems III, A.S.P. Conference Series, 61, Dennis R. Crabtree, R.J. Hanisch, and Jeannette Barnes, eds., p. 292-295.

Zellner, B., D. J. Tholen, E. F. Tedesco 1985. The eight-color asteroid survey - Results for 589 minor planets. *Icarus* **61**, 355-416

Table 1. AMICA Specification

| | | |
|---|---|---|
| AMICA Camera Head | Effective Lens Aperture | 15 mm |
| | Focal Length | 120.80 mm (measurement value) |
| | Field of View | 5.83°×5.69° (measurement value) |
| | CCD Format | 1024×1000 pixels |
| | CCD pixel size | 12 μm square |
| | Pixel Resolution | 20″.490 (measurement value) |
| | Filters<br>(Turnover Rate) | 7 narrow band and 1 wide band<br>(4.69 sec./filter) |
| | Polarizer | 4 position angle glass polarizers |
| | Pixel sampling rate | 3MHz |
| | Gain factor | ~17 DN/$e^-$ |
| | Readout noise | ~60 $e^-$ (measurement value) |
| ONC-AE/E for AMICA control capabilities and specification | A/D conversion | 12 bit |
| | Full-well | 70,000 $e^-$ |
| | Image memory storage | 16 frames |
| | Exposure times | 5.44ms, 8.20ms, 10.9ms, 16.4ms, 21.8ms, 32.8ms, 43.5ms, 65.6ms, 87.0ms, 131ms, 174ms, 262ms, 348ms, 525ms, 696ms, 1.05s, 1.39s, 2.10s, 2.79s, 4.20s, 5.57s, 8.40s, 11.1s, 16.8s, 22.3s, 33.6s, 44.6s, 67.2s, 89.1s, 134s, 178s, <1μs (for smear) |
| | Software pixel binning | None, 2×2, 4×4, 8×8 |
| | Arithmetic operation between 16 image storage | Average<br>Median<br>Mode<br>Sum |
| | Output mode | Lossless image<br>Lossy image (indicated by *Q* factor)<br>Histogram<br>Average value |
| | The other option for analog-to-digital conversion | 12 bit<br>1 bit (for shape model) |

Table 2. Effective wavelength ($\lambda_{eff}$) respect to the Sun and FWHM ($\Delta\lambda_{eff}$) of each filter (nm)

| Filter | ul (u[*1]) UV | b Blue | v Green | w Red | x NIR | p Pyroxene | zs (z[*1]) 1-μm |
|---|---|---|---|---|---|---|---|
| AMICA $\lambda_{eff}$ | 381 | 429 | 553 | 700 | 861 | 960 | 1008 |
| AMICA $\Delta\lambda_{eff}$ | 45 | 108 | 72 | 70 | 81 | 75 | 66 |
| ECAS $\lambda_{eff}$[*1] | 359 | 437 | 550 | 701 | 853 | 948 | 1041 |
| ECAS $\Delta\lambda_{eff}$[*1] | 60 | 90 | 57 | 58 | 81 | 80 | 67 |

[*1] Tedesco et al. (1982)

Table 3. Summary of the AMICA inflight calibration observations

| Date | Objects(spectral type) | Filter(s) | File Name Prefixes | Objectives |
|---|---|---|---|---|
| 11 May 2003 | Undefined | v | ST_0007065288, ST_0009955932 | Function check |
| 14 May 2003 | FF–lamp | b, w | ST_0015411418, ST_0015419833, ST_0015427218 | Flat monitor |
| 22 May 2003 | Undefined | b, w | ST_0037645439, ST_0037652204 | Function check |
| 23 May 2003 | τ Sco (B0)<br>α Sco (M1) | b, w, p | ST_0040316133, ST_0040323910, ST_0040331721, ST_0040339923 | Co-alignment of the NIRS<br>Flux calibration |
| 11 Nov. 2003 | Mars | b,v,w,p | ST_0515848228, ST_0515850188, ST_0515854093, ST_0515857998, ST_0515859942 | Flux check<br>PSF |
| 6 Jan. 2004 | Saturn | ALL | ST_0671283333, ST_0671287205, ST_0671291077, ST_0671294949, ST_0671298821, ST_0671302693 | Flux check<br>PSF |
| | α Aur (G5) | ALL | ST_0670703429, ST_0670707301, ST_0670711173, ST_0670715045, ST_0670718917, ST_0670722789, ST_0670726661 | Flux check<br>PSF |
| | α Ori (M2) | ALL | ST_0670991429, ST_0670995301, ST_0670999173, ST_0671003045, ST_0671006917, ST_0671010789, ST_0671014661 | Flux check<br>PSF |
| 13 Jan. 2004 | β Tau (B7) | ALL | ST_0689973071, ST_0689975015, ST_0689976976, ST_0689978920, ST_0689980881, ST_0689982825, ST_0689984786 | Flux check<br>PSF |
| 7 April 2004 | FF lamp | b,w | ST_0924050357, ST_0924072610, ST_0924082283, ST_0924216172, ST_0924228933 | Flat monitor<br>Dark current |
| 16 May 2004 | Moon | ALL | ST_1032719487, ST_1032725271, ST_1032727705, ST_1032730140, ST_1032732574, ST_1032735009, ST_1032737427, ST_1032739861, | |
| | | | ST_1032742296, ST_1032744730,<br>ST_1032747164, ST_1032749599<br>ST_1032752033, ST_1032754451,<br>ST_1032756886, ST_1033685522,<br>ST_1033688432, ST_1033691359,<br>ST_1033692110, ST_1033697894,<br>ST_1033703694, ST_1033711636 | |
| 17 May 2004 | Moon<br>Earth | ALL | ST_1035428953, ST_1035432827,<br>ST_1035434526, ST_1035436960,<br>ST_1035439395, ST_1035441813,<br>ST_1035444247, ST_1035446682,<br>ST_1035449116, ST_1035451550,<br>ST_1035453985, ST_1035456403,<br>ST_1035458837, ST_1035461272,<br>ST_1035463706, ST_1035466140,<br>ST_1035510876, ST_1035513310,<br>ST_1035515745, ST_1037053655,<br>ST_1037054864, ST_1037056285 | |
| 18, 19 May 2004 | Earth | b, w, p | ST_1038645059, ST_1038648932,<br>ST_1038650975, ST_1038674045,<br>ST_1038697115<br><br>ST_1040207688, ST_1040226918,<br>ST_1040246149, ST_1044538182,<br>ST_1044547266, ST_1044555403,<br>ST_1047283363, ST_1047293672,<br>ST_1047303084 | |
| 1 June 2004 | τ Sco (B0) | ALL | ST_1076259343, ST_1076261287,<br>ST_1076263248, ST_1076265208,<br>ST_1076267153, ST_1076269097,<br>ST_1076271057 | Degradation monitor |
| 26 Apr. 2005 | Taurus region | wide | ST_1986968284 | Distortion<br>Limiting magnitude |
| 29 August, 2004 | α Leo<br>31 Leo,<br>Itokawa | b,v,w,x,p | ST_2332479491, ST_2332498721,<br>ST_2332786728, ST_2332825189,<br>ST_2333933003, ST_2333935192,<br>ST_2333937365, ST_2333939669, | Optical Navigation<br>Calibration |
| | | | ST_2333941809, ST_2333943950, ST_2333946008, ST_2333948132, ST_2333950223, ST_2333953001, ST_2333955746, ST_2333966529, ST_2334384682, ST_2334386774, ST_2334388930, ST_2334391087 | |
| 28 Sep. 2005 | α Crv (G5)<br>β Crv (F0) | b,v,w,x,p | ST_2416006923, ST_2416009832, ST_2416012758, ST_2416015651, ST_2416018577, ST_2416021503, ST_2416024429, ST_2416027355, ST_2416030248, ST_2416033157, ST_2416036083, ST_2416038993, ST_2416041902, ST_2416044828, ST_2416047721, ST_2416050631, ST_2416053557, ST_2416056450, ST_2416059376, ST_2416062302, ST_2416065195, ST_2416068104, ST_2416071030, ST_2416073923, ST_2416076849, ST_2416079775, ST_2416082668, ST_2416085578, ST_2416088504, ST_2416091397 | PSF Calibration |
| 30 October, 2005 | α Vir (B1) | ALL | ST_2503211819, ST_2503219515, ST_2503227228, ST_2503234956, ST_2503242652, ST_2503250381, ST_2503258093 | PSF Calibration |
| | FF lamp | b, w | ST_2503053162, ST_2503054143, ST_2503267652, ST_2503279188, ST_2503290757, ST_2503302309 | |

Table 4. Geometric calibration Table (Eq. (2)–(3))

| Date (filename) | N[*1] | $\alpha_0$ | $\delta_0$ | $\theta$ (degree) | $K_1$ | $\Delta_{\mathrm{PIX}}$ (arcseconds) | $F$ (mm) | Ave error (pixels) |
|---|---|---|---|---|---|---|---|---|
| 2005/04/26 (ST_2313701479) | 62 | $3^h55^m25^s.8$ | +20°42′31″ | 257.34 | 0 | 20.490±0.005 | 120.80±0.03 | 0.4 |
| | | | | | $-2.8\times10^{-5}$ | 20.505±0.005 | 120.71±0.03 | 0.3 |

[*1] Number of stars used to examine the geometric calibration

Table 5. AMICA Pre−flight Dark Model Parameters (Eq. (4)–(5))

| | | |
|---|---|---|
| Dark model (imaging area) | $d_{0,\mathrm{IM}}$ | $d_{1,\mathrm{IM}}$ |
| | 3.60 | 0.123 |
| Dark model (store area) | $d_{0,\mathrm{ST}}$ | $d_{1,\mathrm{ST}}$ |
| | 2.96 | 0.140 |

Table 6. Mission Phase Hot Pixel List (>250 DN $sec^{-1}$)

| H | V | Dark rate (DN $sec^{-1}$) |
|---|---|---|
| 407 | 300 | 540 |
| 599 | 408 | 330 |
| 820 | 14 | 310 |
| 930 | 624 | 305 |
| 897 | 716 | 290 |

Table 7. AMICA Inflight Bias Model Parameters (Eq. (6))

| Bias model parameters | $B_0$ | $B_1$ | $B_2$ |
|---|---|---|---|
| | 3.18x10$^{2}$ | -4.12x10-$^{2}$ | 2.00x10$^{-5}$ |

Table 8. List of stars observed by AMICA in v-band

| | Spectral type[*1] | Observed date (phase [*2]) | V mag [*1] | B-V [*1] | $V_{MAX}$ / $V_{MIN}$ [*3 4] |
|---|---|---|---|---|---|
| τ Sco | B0 | 2003 / 05 / 23 (IOP)<br>2004 / 06 / 01 (CP) | 2.82 | -0.25 | — |
| α Aur | G5 | 2004 / 01 / 06 (CP) | 0.08 | 0.80 | — |
| α Ori[*5] | M2 | 2004 / 01 / 06 (CP) | 0.50 | 1.85 | 0.42 / 1.30 [*3] |
| β Tau | B7 | 2004 / 01 / 13 (CP) | 1.65 | -0.13 | — |
| α Leo | B7 | 2005 / 08 / 29 (AP) | 1.35 | -0.11 | 1.33 / 1.40 [*4] |
| 31 Leo | K4 | 2005 / 08 / 29 (AP) | 4.37 | 1.45 | — |
| α Crv | F0 | 2005 / 09 / 28 (MP) | 4.02 | 0.32 | — |
| β Crv | G5 | 2005 / 09 / 28 (MP) | 2.65 | 0.89 | 2.60 / 2.66 [*4] |
| α Vir | B1 | 2005 / 10 / 30 (MP) | 0.98 | -0.23 | 0.97 / 1.04 [*3] |

[*1] Spectral type, V-band magnitude and B-V color index in Bright Star Catalog 5th Edition (Hoffleit and Jaschek 1991).

[*2] The definition of Mission phases are summarized in Section 3. IOP: Initial operation phase, CP: Cruise phase, AP: Approach phase, MP: Mission phase.

[*3] Maximum and minimum magnitude listed in General catalogue of Variable Stars 4th ed. (Kholopov 1996).

[*4] Maximum and minimum magnitude listed in the New Catalogue of Suspected Variable Stars Supplement – Version 1.0 (Kazarovets et al. 1998)

[*5] We excluded the obtained instrumental magnitude of this star from Fig. 16 because of the large variability.

Table 9. Conversion factor from DN into Radiance or Reflectivity

| Filter | $\lambda_{eff}$ (μm) | Conversion into Radiance (W $m^{-2}$ $\mu m^{-1}$ sr-$^{1}$) / (DN sec-1) | Scale factor (error) |
|---|---|---|---|
| ul | 381 | — | 6.259 (0.063) [*2] |
| b | 429 | — | 1.254 (0.008) |
| v | 553 | $3.42\times10^{-3}$ | 1 |
| w | 700 | — | 0.645 (0.005) |
| x | 861 | — | 0.600 (0.006) |
| p | 960 | ($1.89\times10^{-3}$) [*1] | 1.514 (0.014) |
| zs | 1008 | — | —[*3] |

[*1] Nakamura et al., 2005,

[*2] We regarded AMICA ul-band as the standard ECAS u-band, and compared AMICA intensity with the ground-based observation results by UH88.

[*2] No available data were obtained from the ground.

## Figure Captions

Figure 1.

The bottom view of the Hayabusa spacecraft. AMICA is installed on the bottom ($-Z_{SC}$) plane. In the normal attitude, the Sun and the Earth are located in the $+Z_{SC}$ direction, and Itokawa in the $-Z_{SC}$.

Figure 2.

Design of AMICA imaging section. There are two masked areas on the right and left edge. AMICA has four position angle polarizers, which comprise 200 x 200 pixels each and are located in on the left edge of the imaging section. The horizontal and vertical axes are designated by "H" and "Y" starting from 0 coordinate value. The spacecraft coordinates are indicated by solid arrows and the direction of the position of FF–lamps are given by double-lined arrows.

Figure 3.

The transmittance (top) and the system efficiency (bottom) of AMICA seven band filter. The filter system is nearly equivalent to the ECAS system.

Figure 4.

AMICA Front view.

Figure 5.

Schematic illustration of the AMICA in-flight calibration pipeline.

Figure 6.

Linear-log plots of dark current in (a) imaging area and (b) masked area as a function of CCD temperature. The dashed lines are the fit to the data using Eq. (4) and (5) and the coefficients are listed in Table 5. Gray area corresponds the operation temperature after the launch.

Figure 7.

Temperature from telemetry data on AMICA CCD for all images obtained after the launch. Temperatures stayed between -30 and -24 degree Celsius before the sampling phase, and it increased up to -19 degree Celsius during the touchdown rehearsal on November 12.

Figure 8.

Histogram of dark pixels in Initial Operation Phase (solid line), Cruise Phase (dashed line) and Approach Phase (dotted line).

Figure 9.

History of CCD bias level obtained from short time exposure (<10 sec) of dark sky. The dark current was subtracted from the original data.

Figure 10.

w-band flatfield images. Left and center images were obtained with the NT Space integrating sphere, and the right image was obtained by a portable integrating sphere. The dark doughnut-shaped pattern, indicated by the white arrows, was caused by the nonuniformity in the integrating sphere.

Figure 11.

Example of w-band (a) pre–flight flatfield image, (b) pre–flight FF–lamp image, (c) inflight FF–lamp image taken in Initial Operation Phase and (d) Cruise Phase. Images of (e) and (f) are the ratio of (c) to (b) and (d) to (c), respectively. The contour levels in (e) and (f) are 1% of the mean intensity.

Figure 12.

Temporal variations in the FF–lamp images during the Cruise phase. The vertical axis represents the average in the H = 462–561 and V = 462–561 square normalized at May 2003.

Figure 13.

(a) The effects of readout smear removal by accident to $Y_{SC}$-reaction wheel. The left image was taken on 14 October 2005 (ST_2459278619) ns the right image is the same as the right image but contrast-enhanced one. The inadequate smear correction generated the vertical streaks indicated by arrows. (b) The effects of readout smear removal in a w-band image. (Left) Raw image. (Center) Raw image with the contrast enhanced to bring out the smear. (Right) Contrast–enhanced image after smear removal. The residuals are below 1 DN.

Figure 14.

Linearity plots from preflight data measured at room temperature.

Figure 15.

Linearity of AMICA CCD examined from Itokawa images taken at different exposure times. (a) An image taken at exposure time of 131.0 ms. Two boxes are indicated the positions of brighter terrain and darker train.(b) Results are given as the rate of signal accumulation per unit time at each exposure time. These data were obtained on 21 September 2005.

Figure 16.

Comparison between Johnson $V_{CAT}$ magnitude and instrumental magnitude $Vi_{ns}$t (Eq. (9)) for the stars observed inflight.

Figure 17.

Comparison between spectra taken at different rotational phase (a) and different times of day (b).

We also show the spectra taken by the ground-based telescopes.

Figure 18.

Plot of the point spread functions in (a) ul, (b) b, (c) v, (d) w, (e) x, (f) p, and (g) zs band filters. Large fluctuations at large distances are due to the noise.

Figure 19.

Normalized intensity of scattered light from the limb of the moon: (a) linear–linear and (b) log-linear.

Figure 20.

Earth image: (top) after flat, bias and dark current removal and (bottom) the image restored by RL algorithm . (ST_1038650975 — ST_1038697115)

Figure 21.

Definitions of the azimuth angle and the tilt angle. The azimuth angle, $\phi$, is the angle between the $+X_{SC}$ direction of the spacecraft and the Sun-pointing vector projected on the spacecraft $X_{SC}$–$Y_{SC}$ plane. $\theta$ is the tilt angle of the $Z_{SC}$–$Y_{SC}$ plane. The tilt angle was limited to within 10˚.

Figure 22.

Stray light (sky flux – dark current) rate versus spacecraft azimuth angle $\phi$. The nominal attitude of Hayabusa corresponds to the zero azimuth angle, where the Sun illuminates only the +X panel. The large azimuth angle possibly allows for stray light from the direct illumination of the nearby sampler horn.

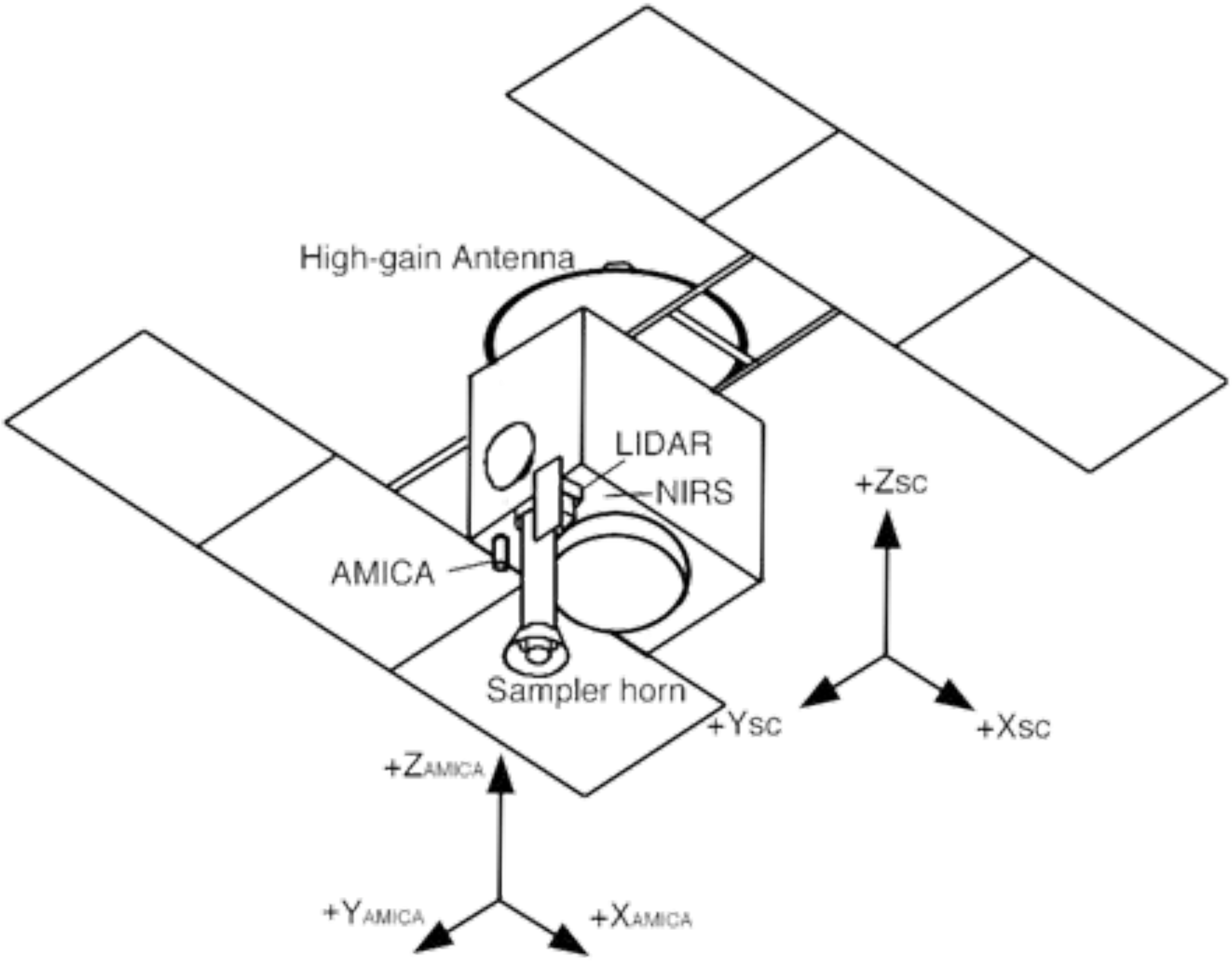

Figure 1

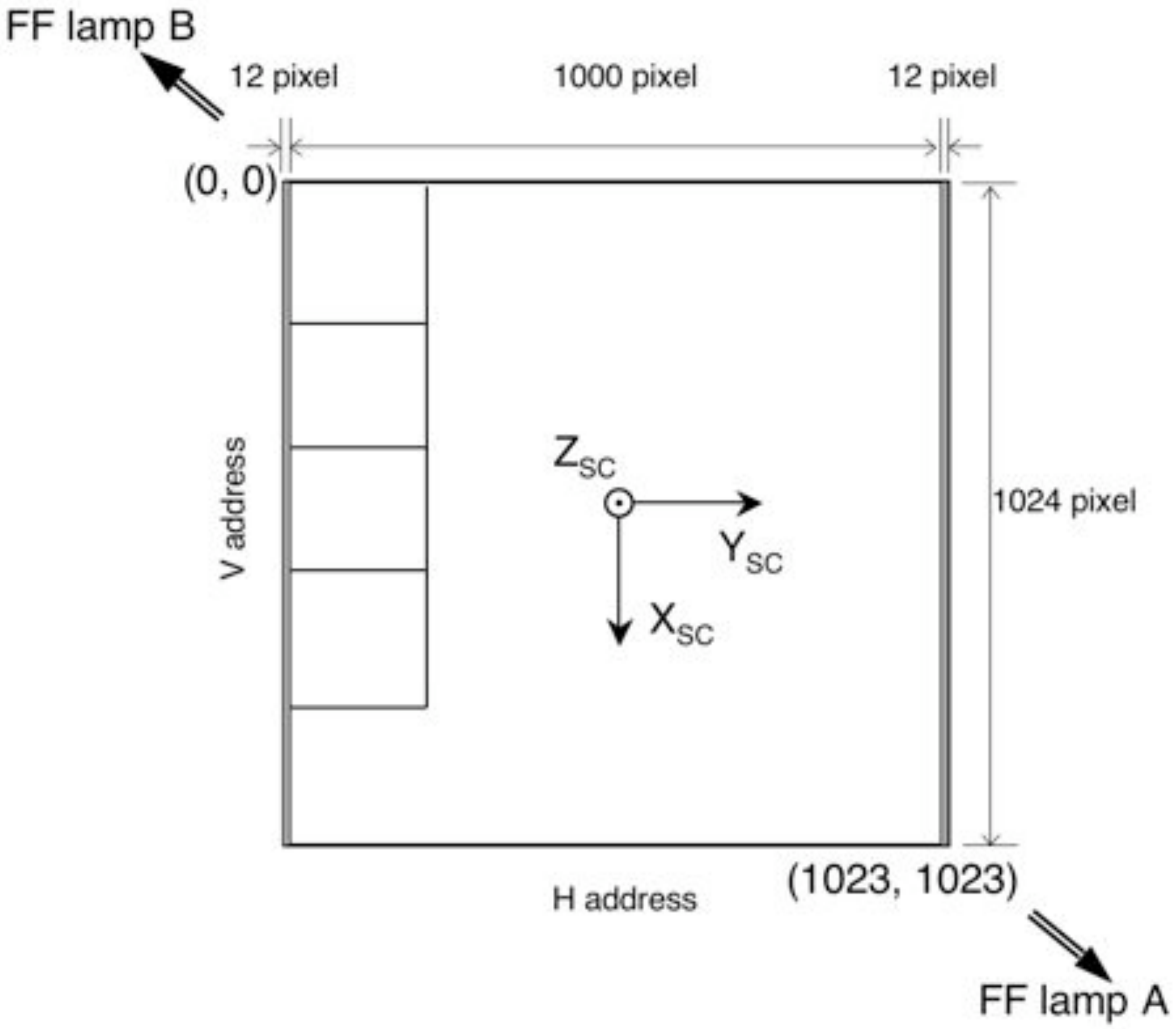

Figure 2.

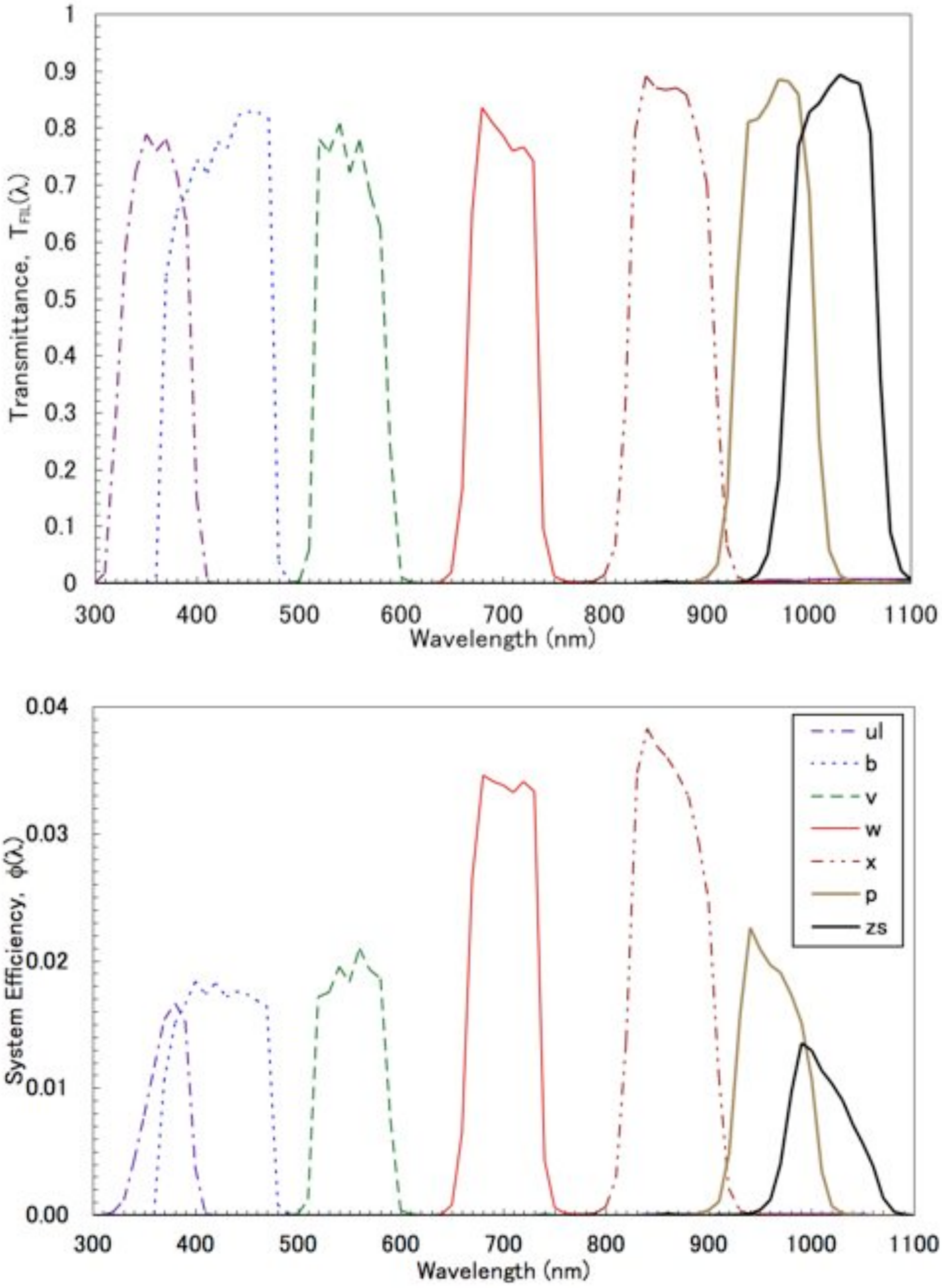

Figure 3

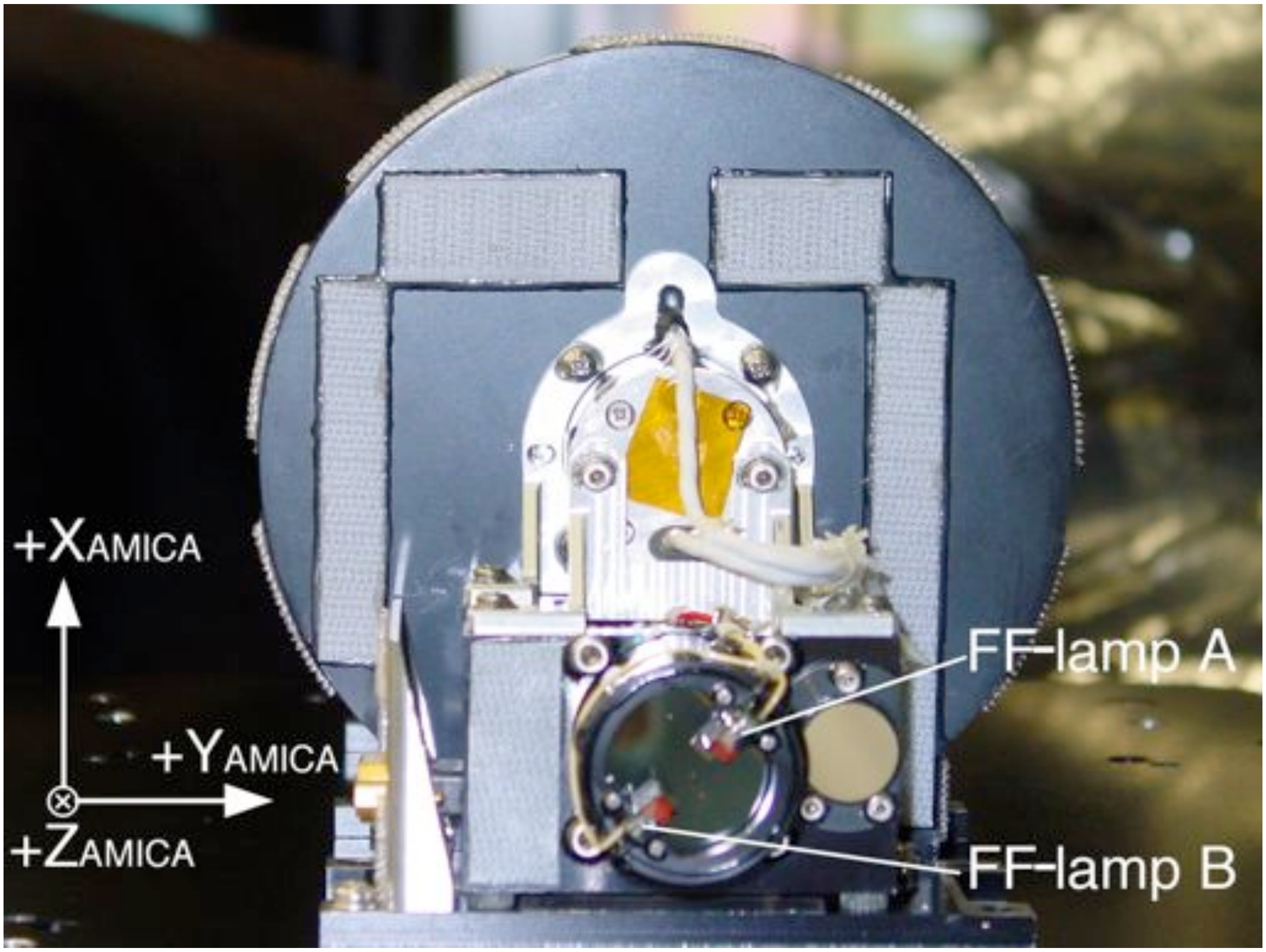

Figure 4.

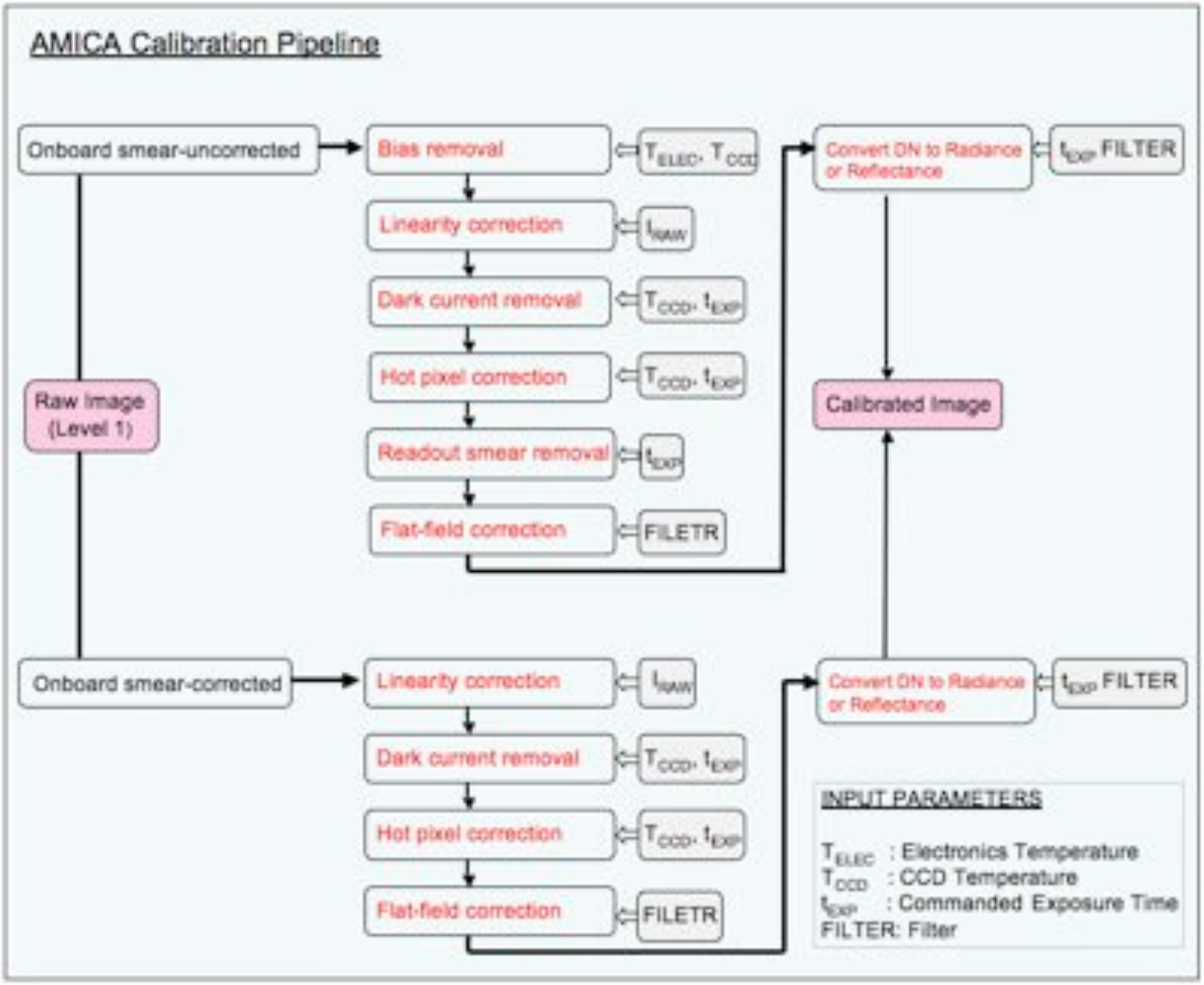

Figure 5.

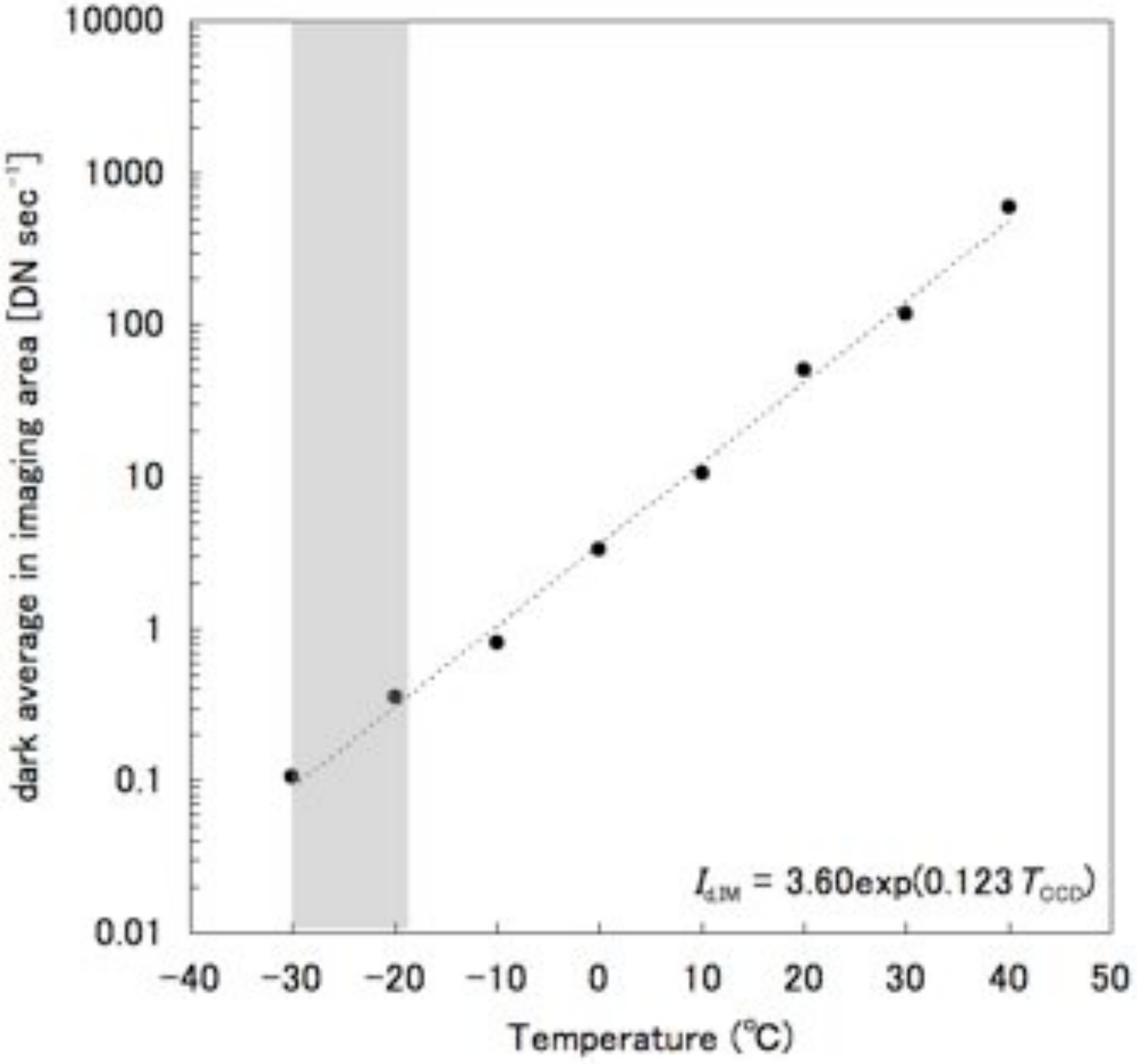
10000
1000
100
10
1
0.1
0.01
dark average in imaging area [DN sec⁻¹]
$I_{d,IM}$ = 3.60exp(0.123 $T_{CCD}$)
−40 −30 −20 −10 0 10 20 30 40 50
Temperature (°C)

Figure 6 (a).

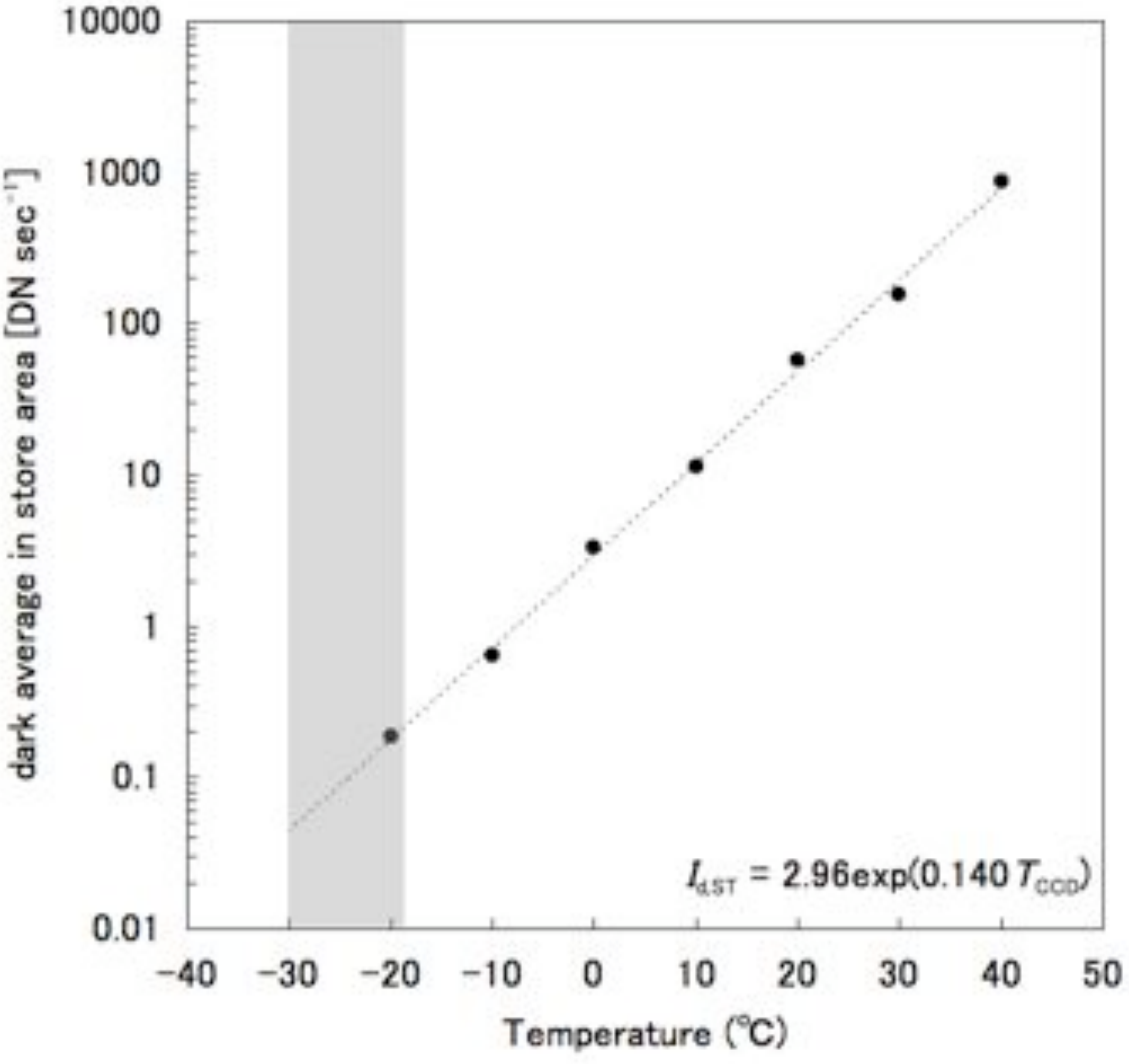

Figure 6 (b).

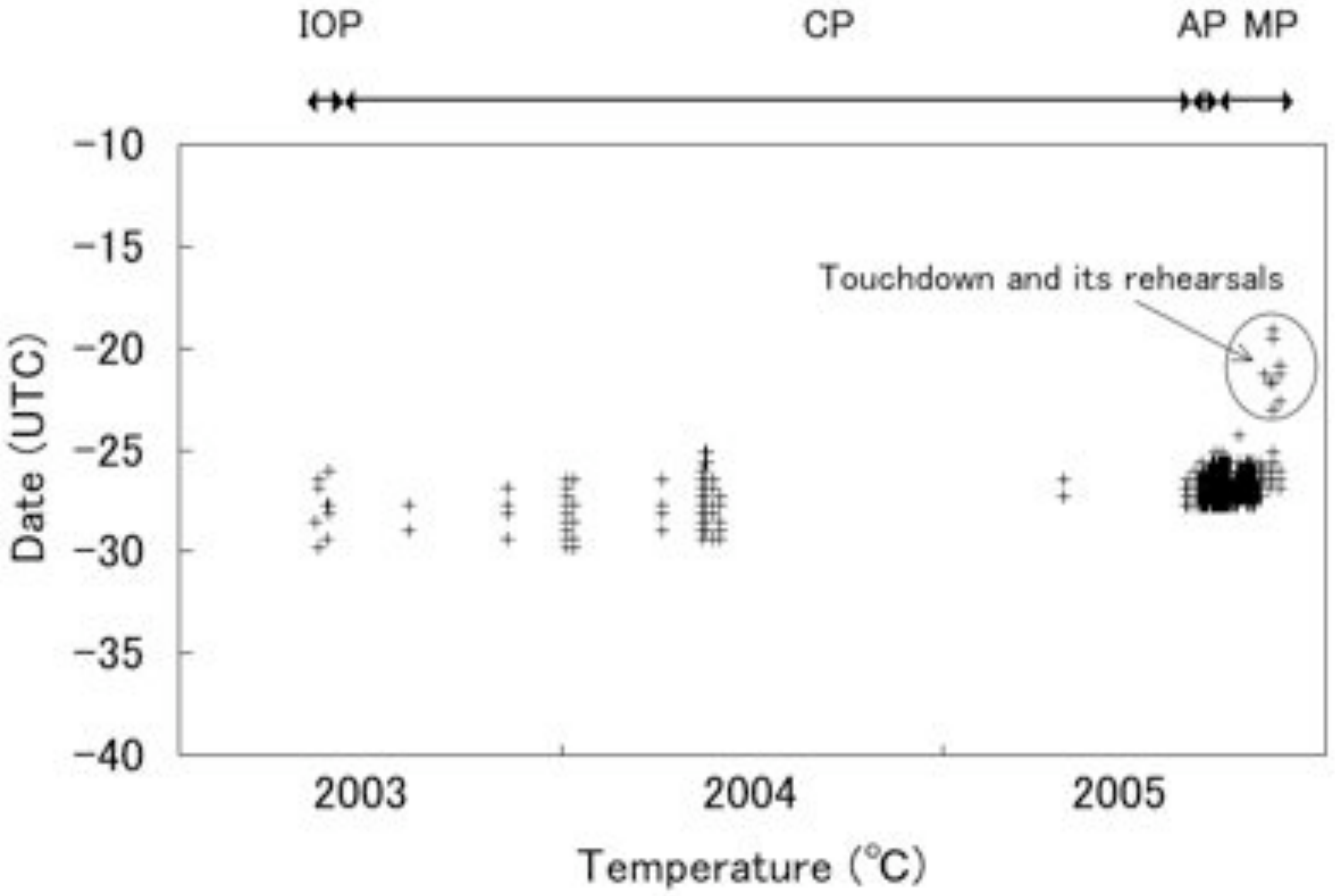

Figure 7. Temperature history.

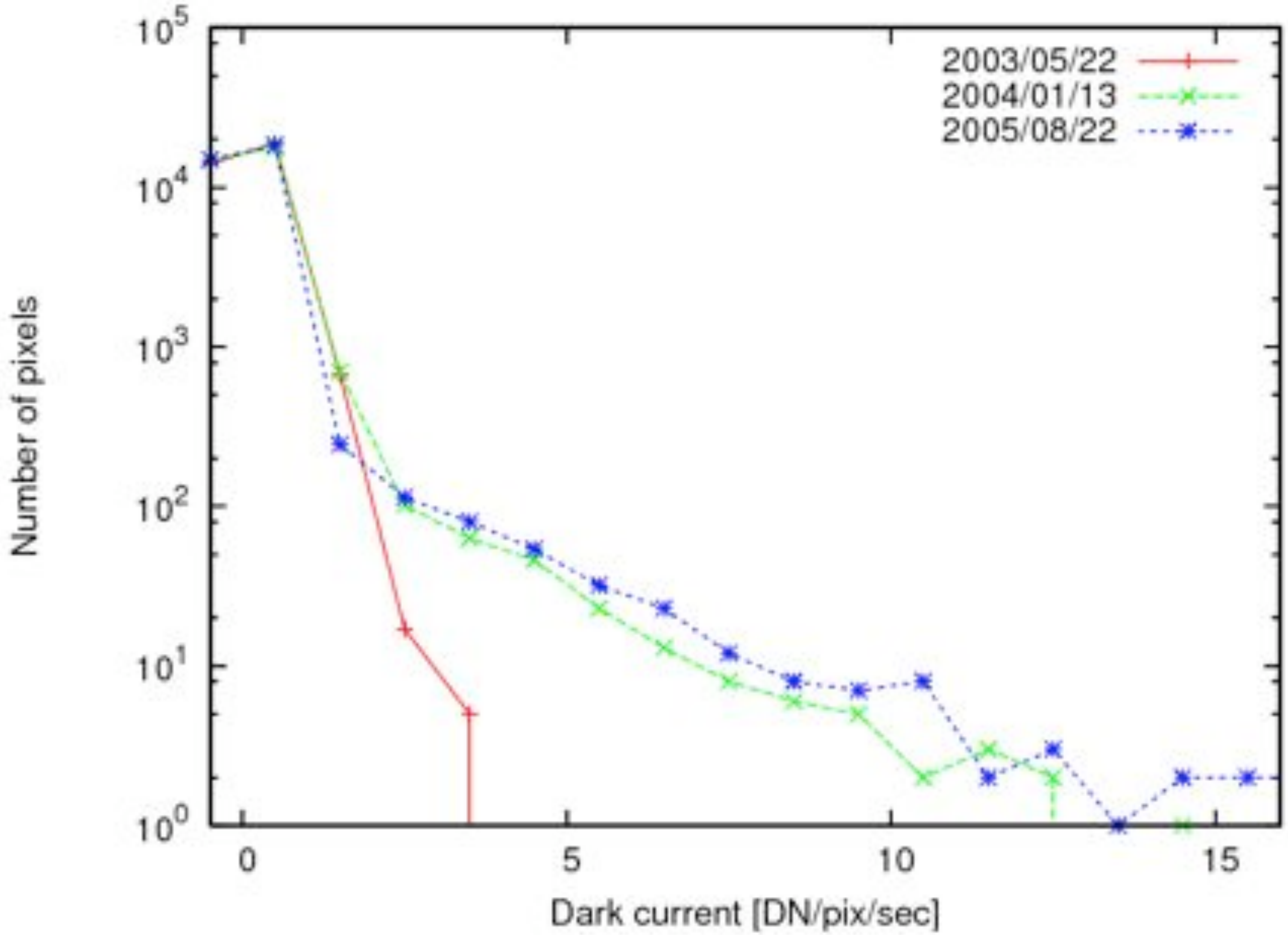

Figure 8. Distribution of dark counts.

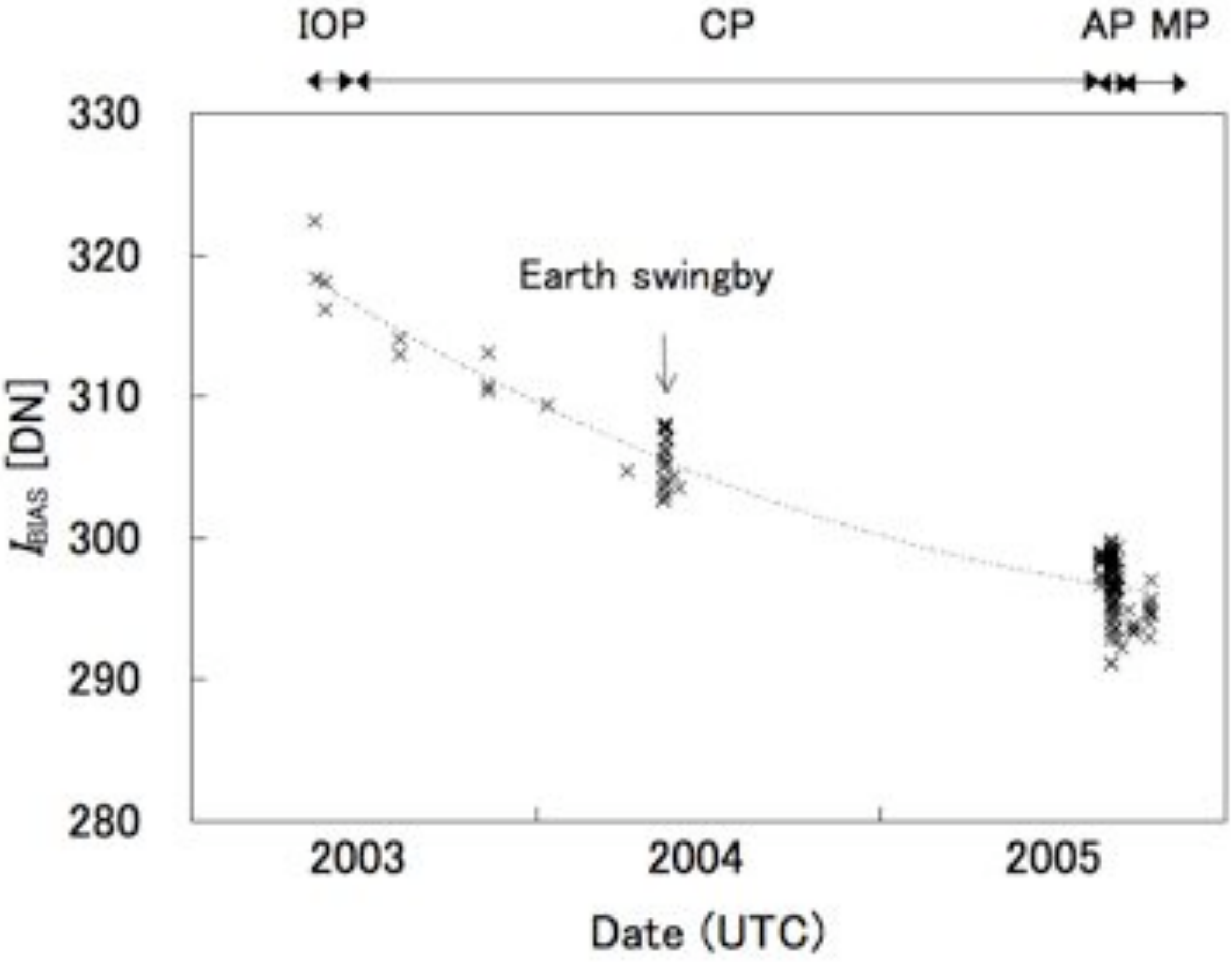

Figure 9.

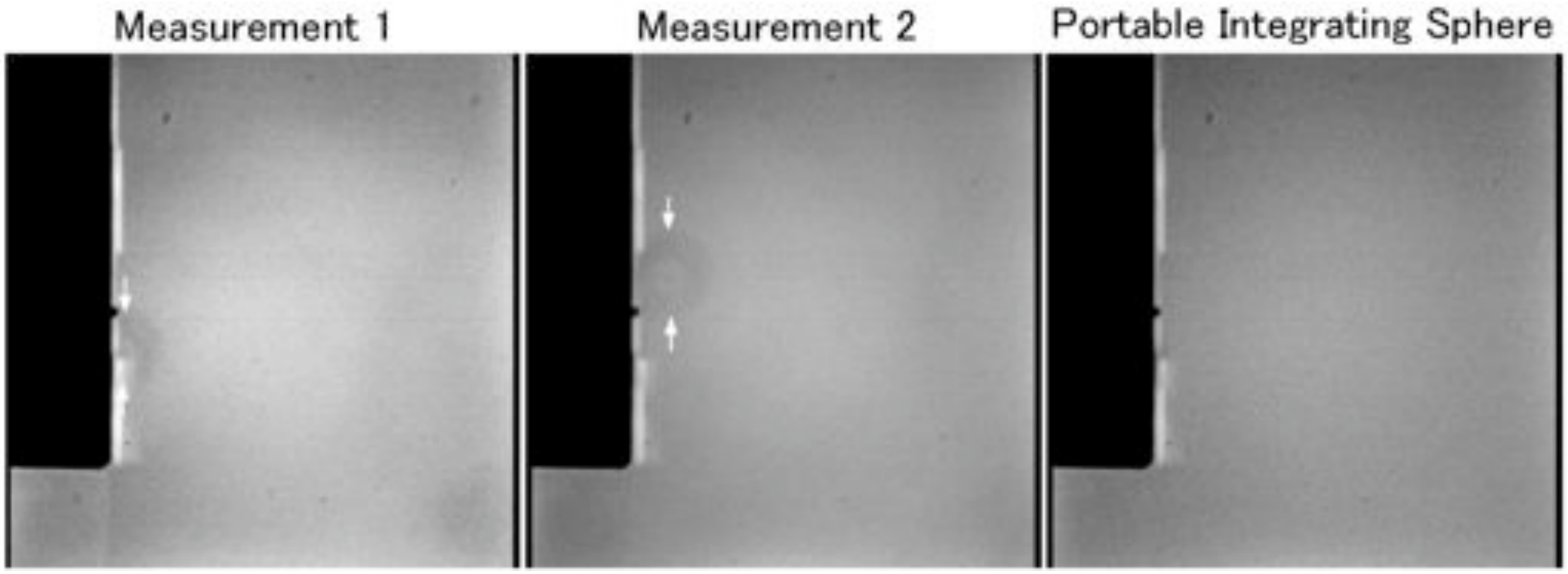

Figure 10.

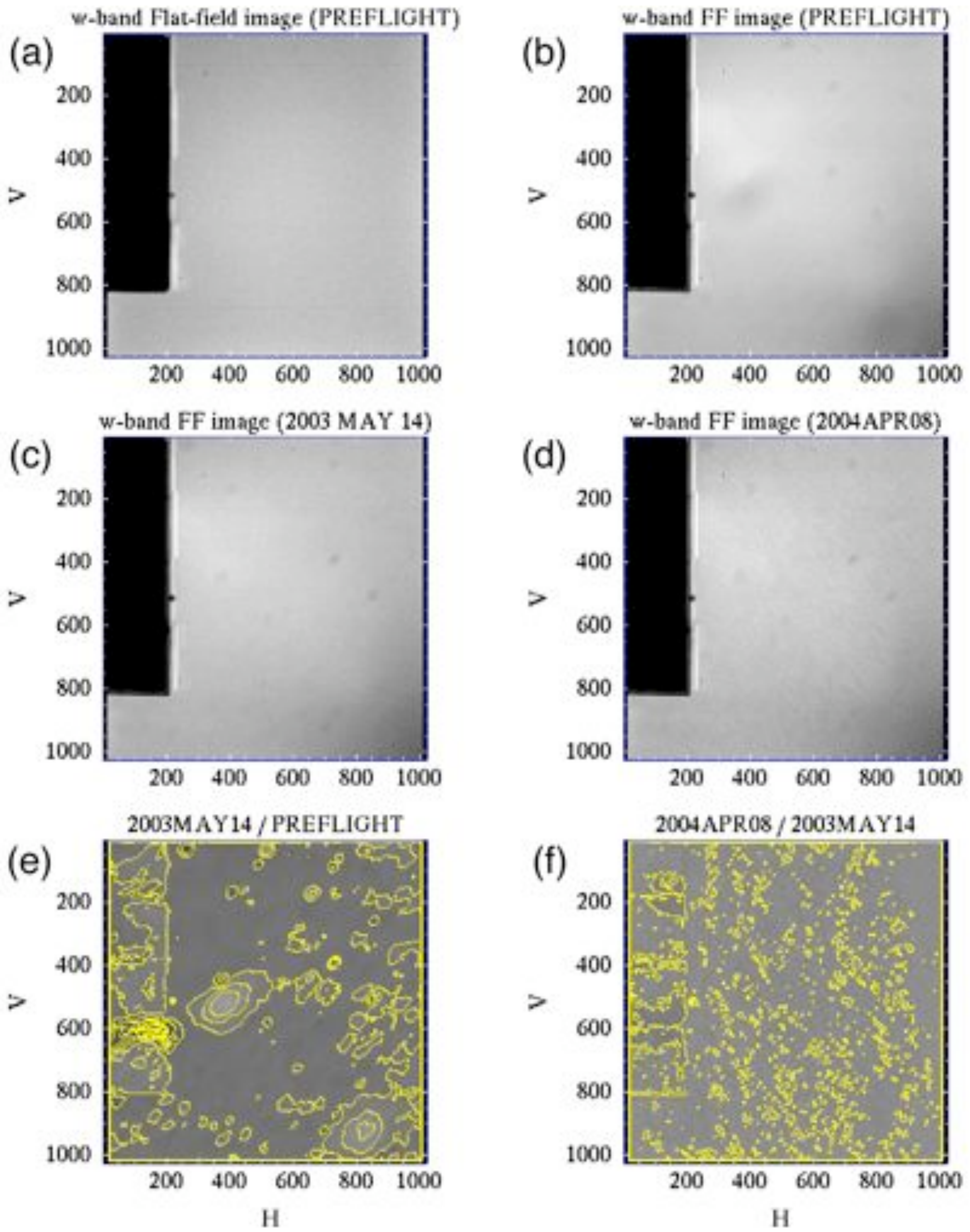

Figure 11. Flat-field and FF-lamp images

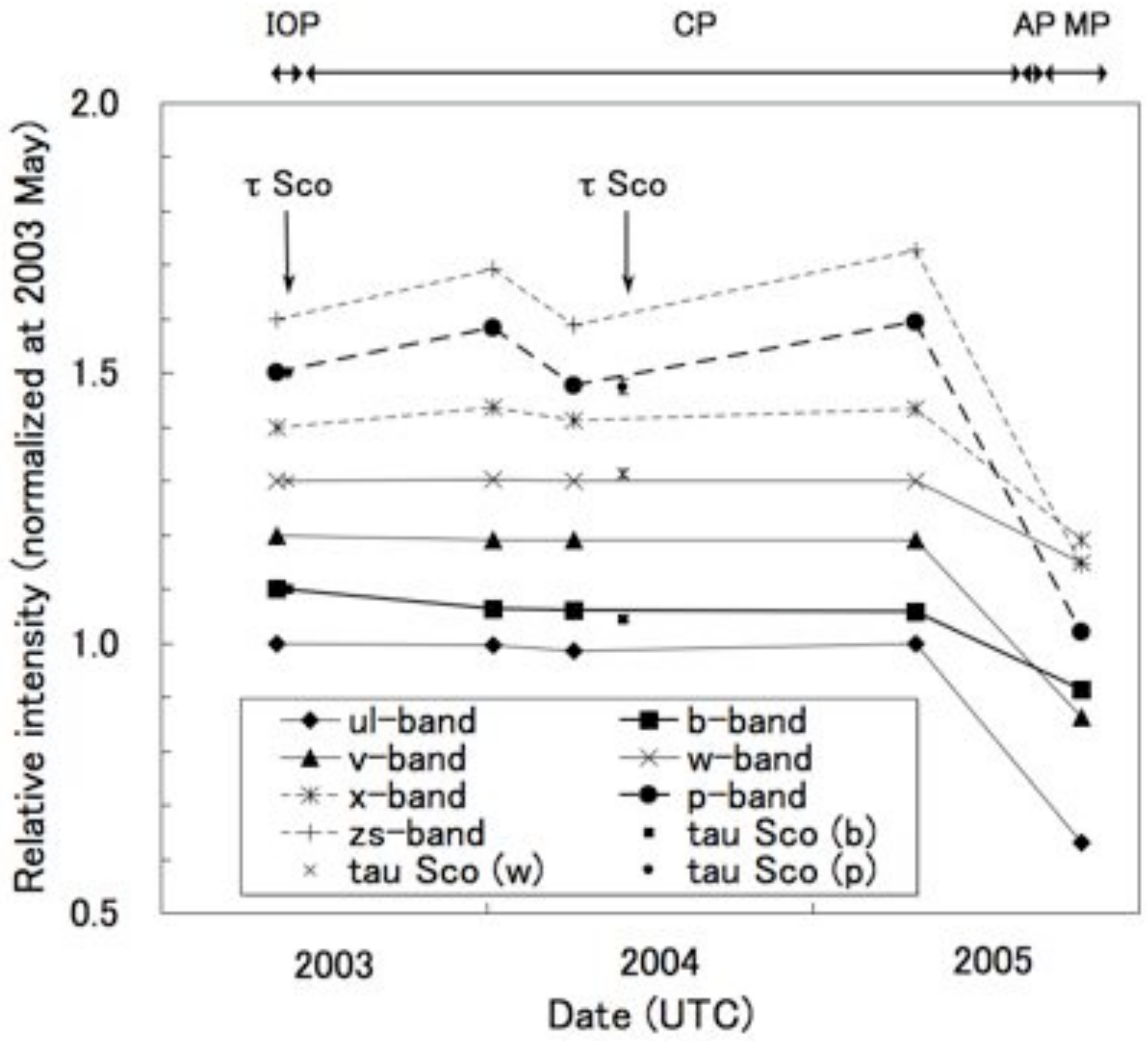

Figure 12. Degradation monitor.

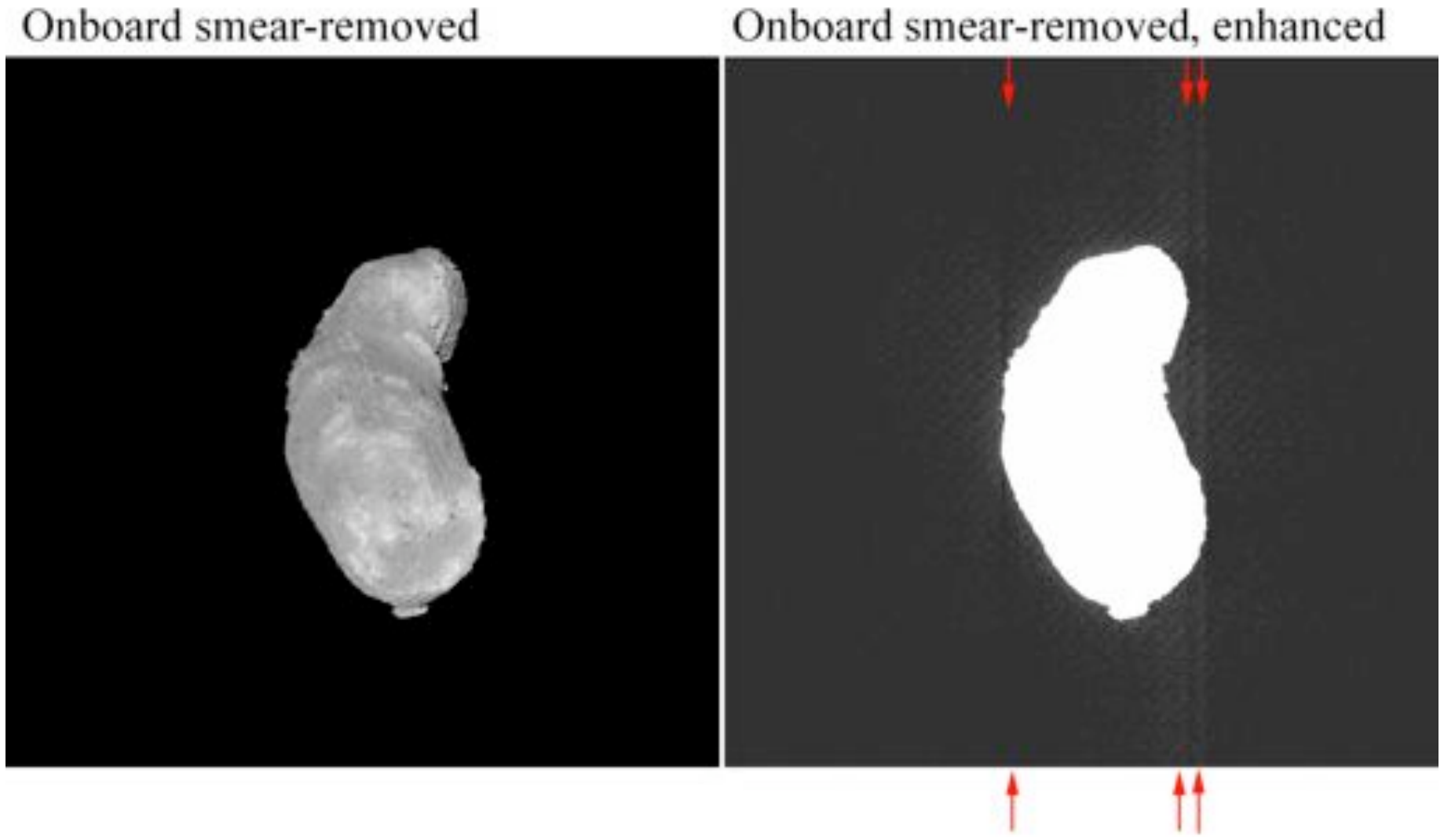

Figure 13(a)

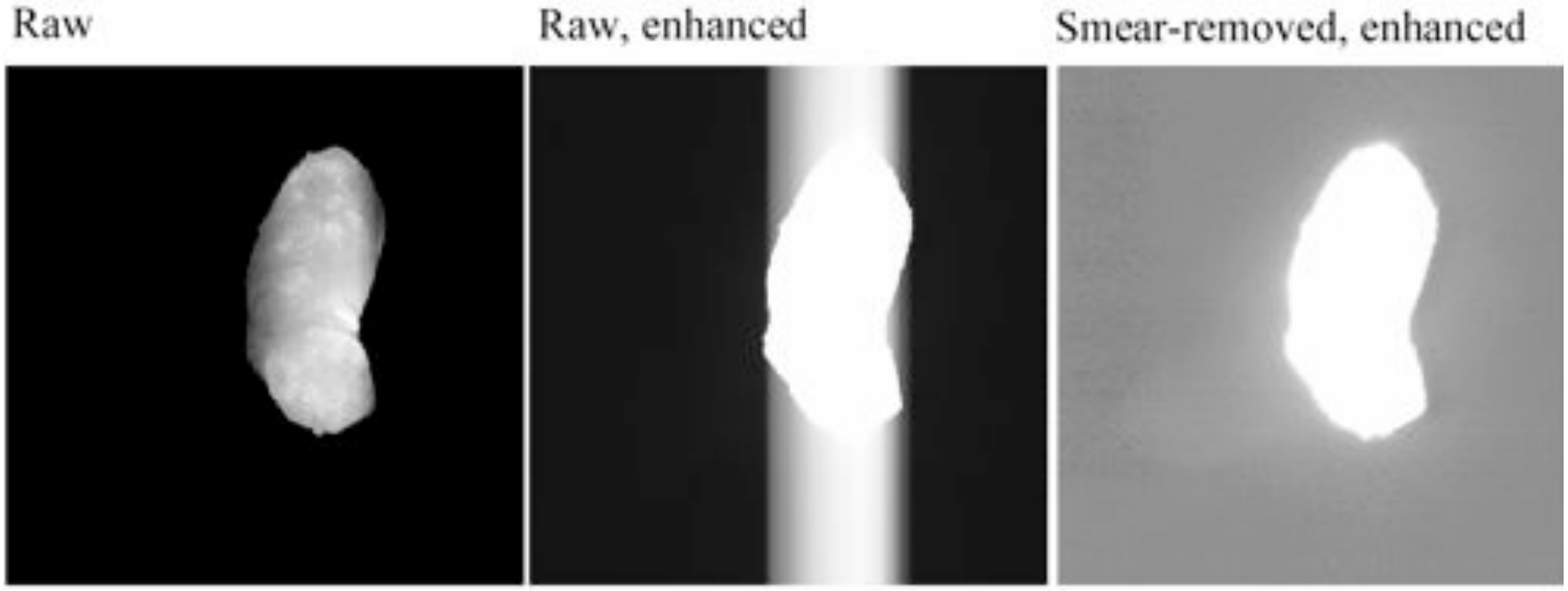

Figure 13 (b)

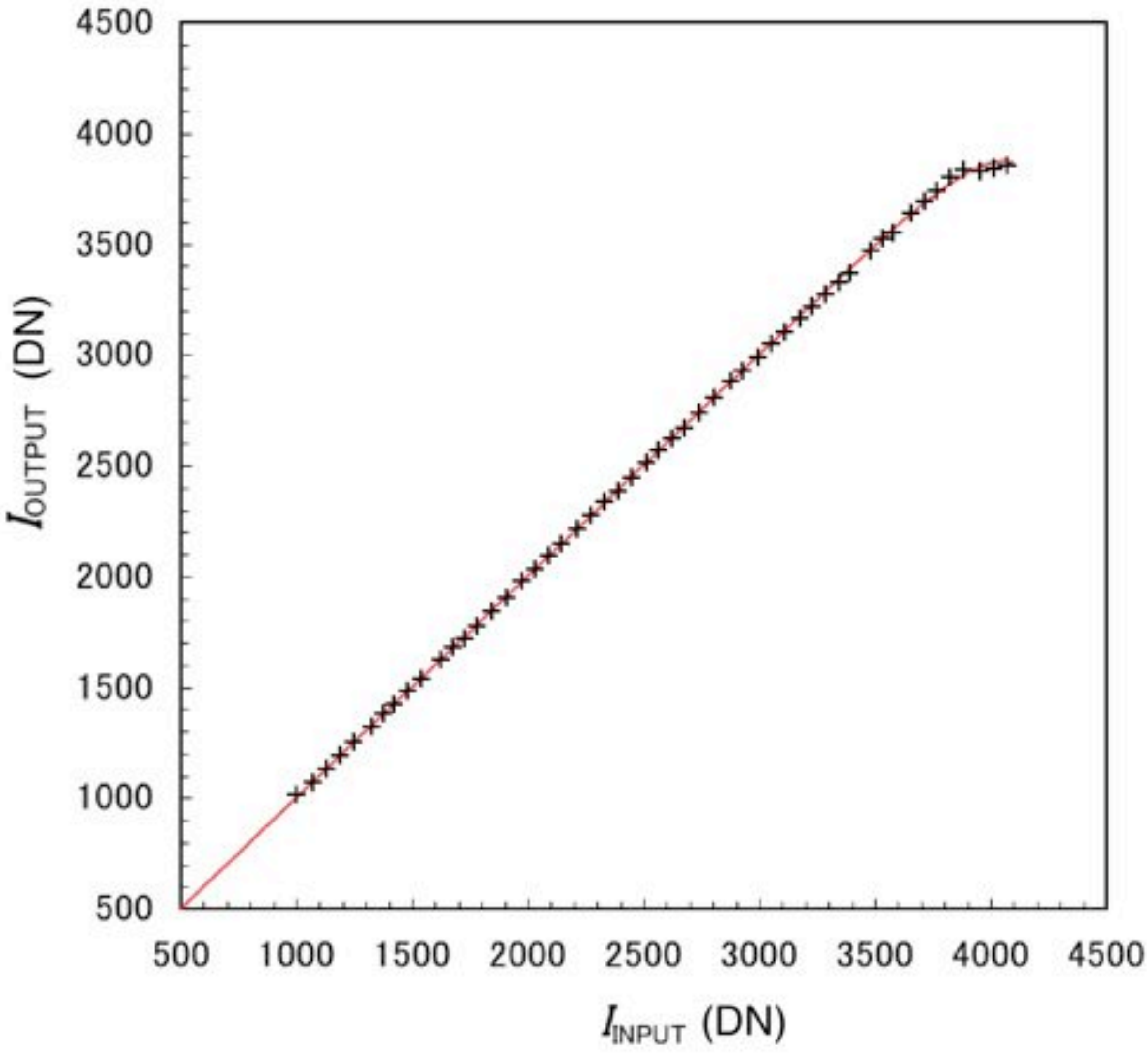

Figure 14.

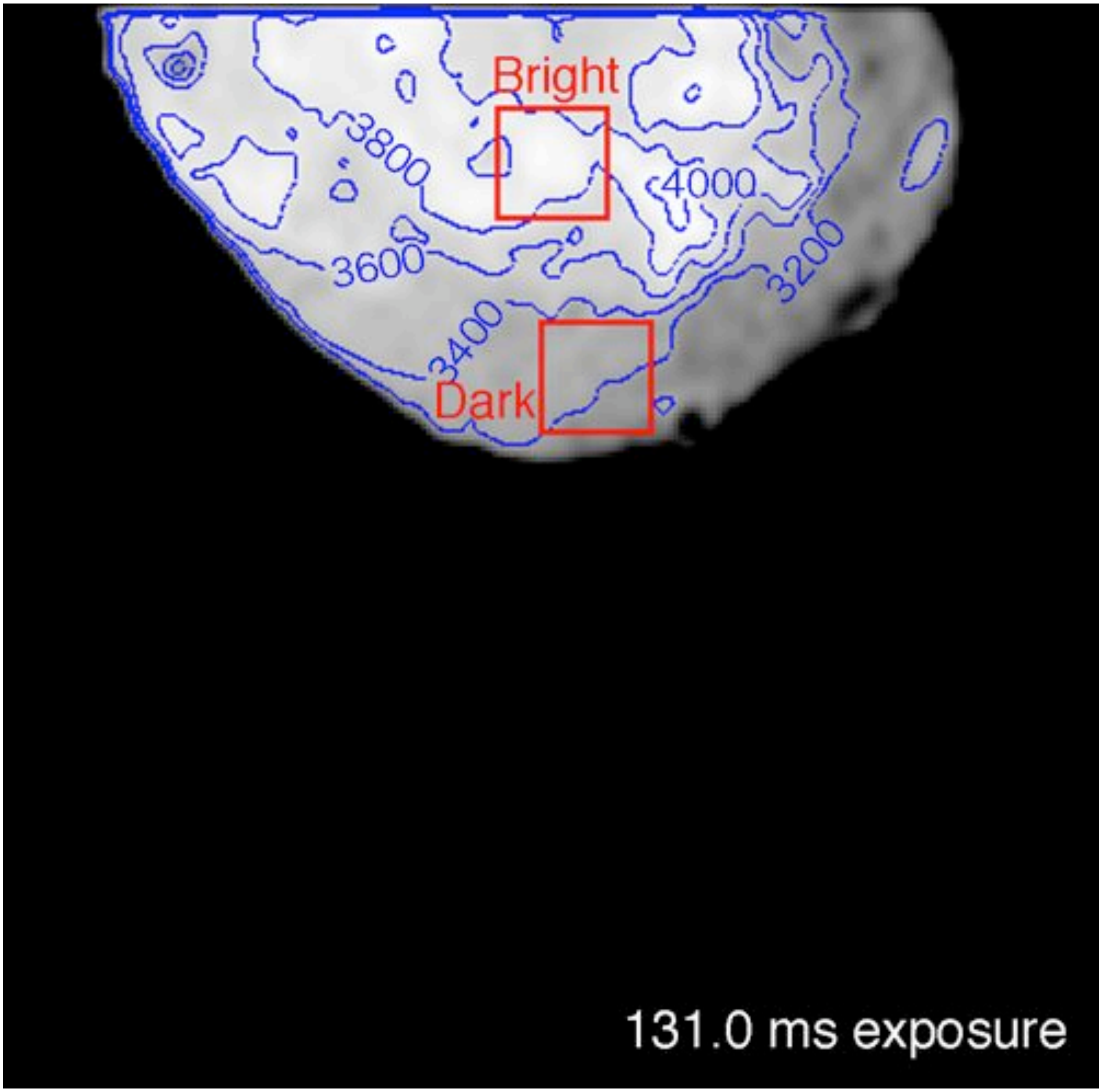

Figure 15(a)

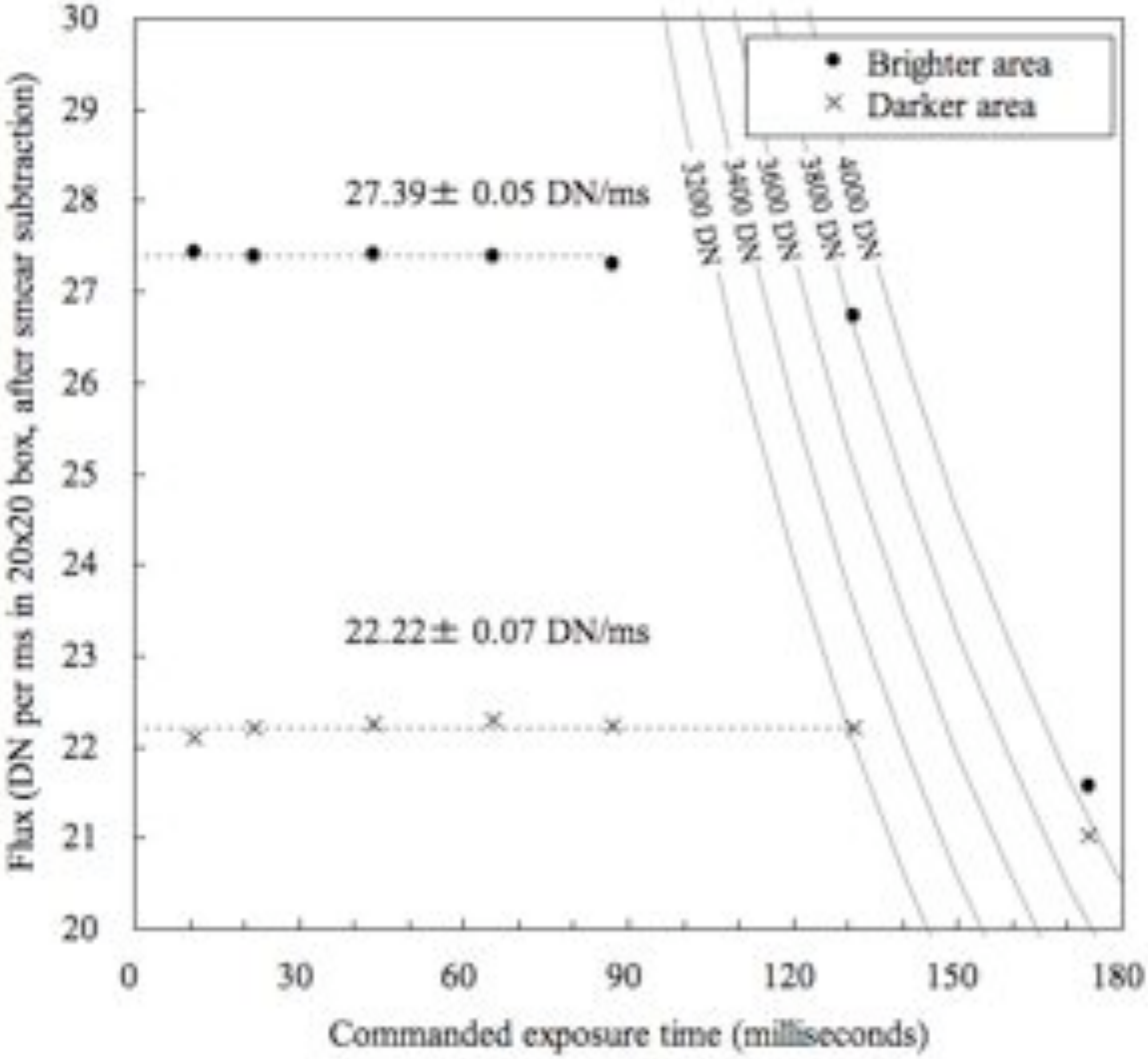
Brighter area
Darker area
27.39± 0.05 DN/ms
22.22± 0.07 DN/ms
3200 DN
3400 DN
3600 DN
3800 DN
4000 DN
Flux (DN per ms in 20x20 box, after smear subtraction)
Commanded exposure time (milliseconds)
30
29
28
27
26
25
24
23
22
21
20
0
30
60
90
120
150
180

Figure 15 (b)

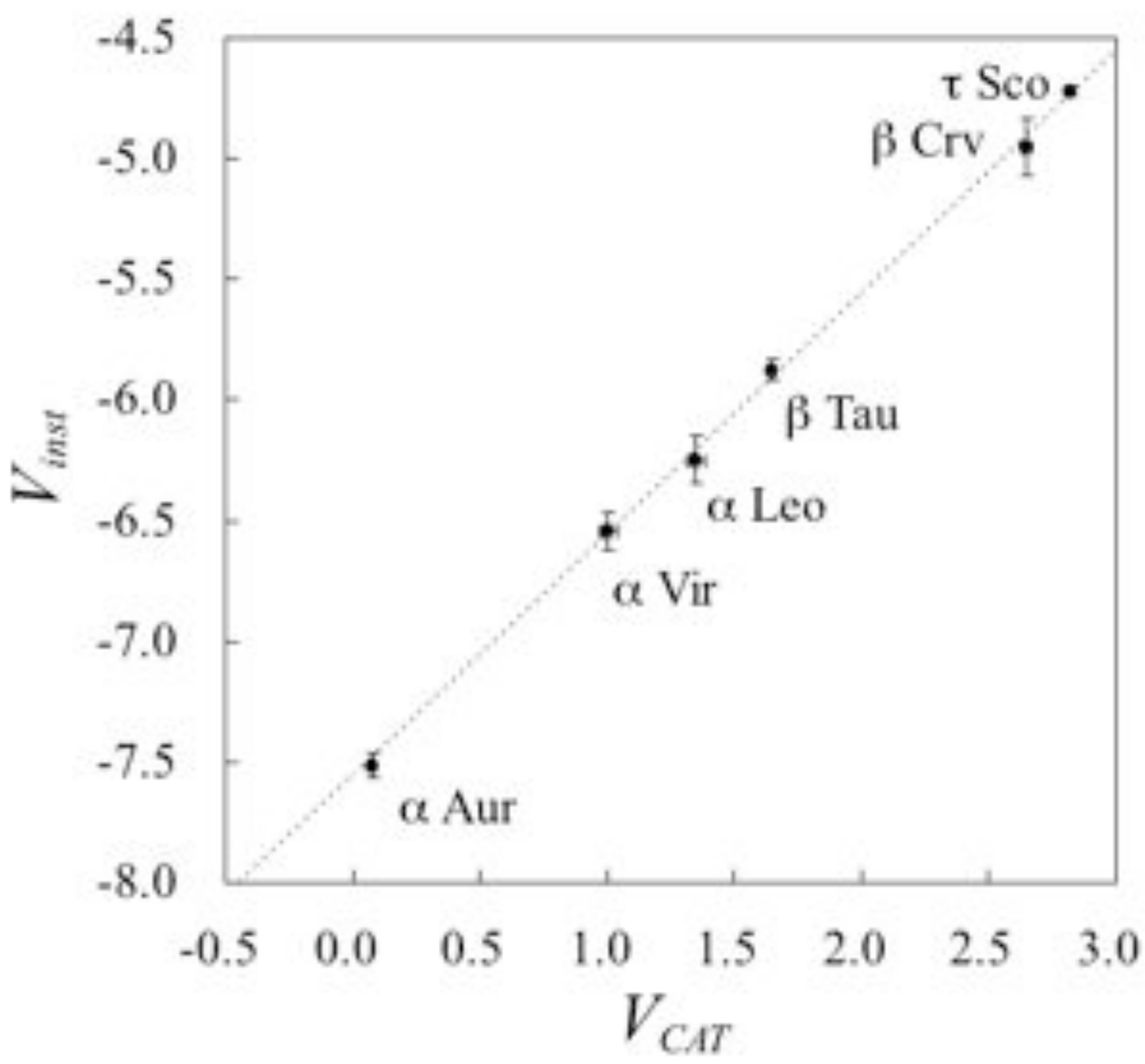

Figure 16.

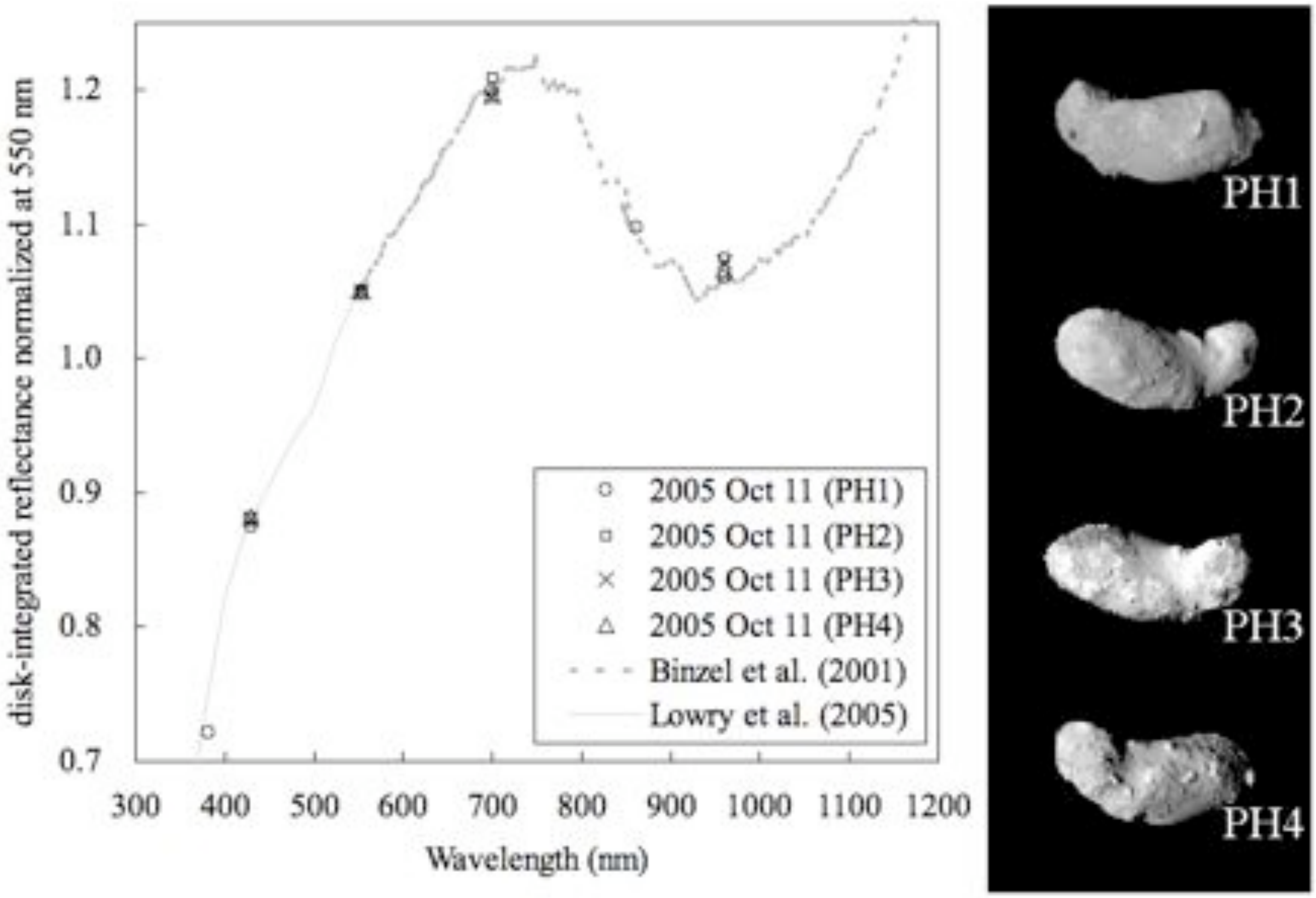
1.2
1.1
1.0
0.9
0.8
0.7
disk-integrated reflectance normalized at 550 nm
300 400 500 600 700 800 900 1000 1100 1200
Wavelength (nm)
2005 Oct 11 (PH1)
2005 Oct 11 (PH2)
2005 Oct 11 (PH3)
2005 Oct 11 (PH4)
Binzel et al. (2001)
Lowry et al. (2005)
PH1
PH2
PH3
PH4

Figure 17. (a)

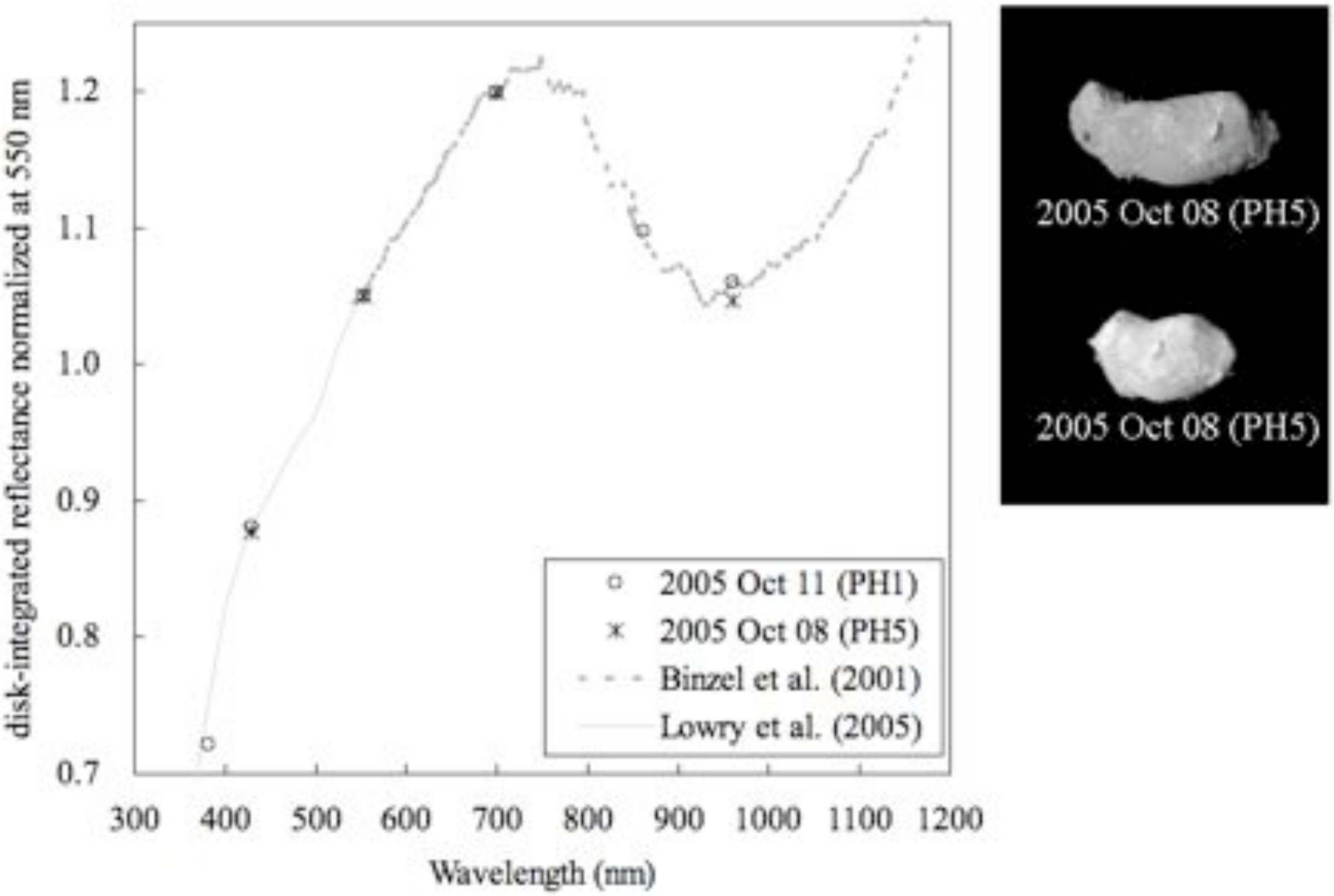
1.2
1.1
1.0
0.9
0.8
0.7
disk-integrated reflectance normalized at 550 nm
300
400
500
600
700
800
900
1000
1100
1200
Wavelength (nm)
2005 Oct 11 (PH1)
2005 Oct 08 (PH5)
Binzel et al. (2001)
Lowry et al. (2005)
2005 Oct 08 (PH5)
2005 Oct 08 (PH5)

Figure 17. (b)

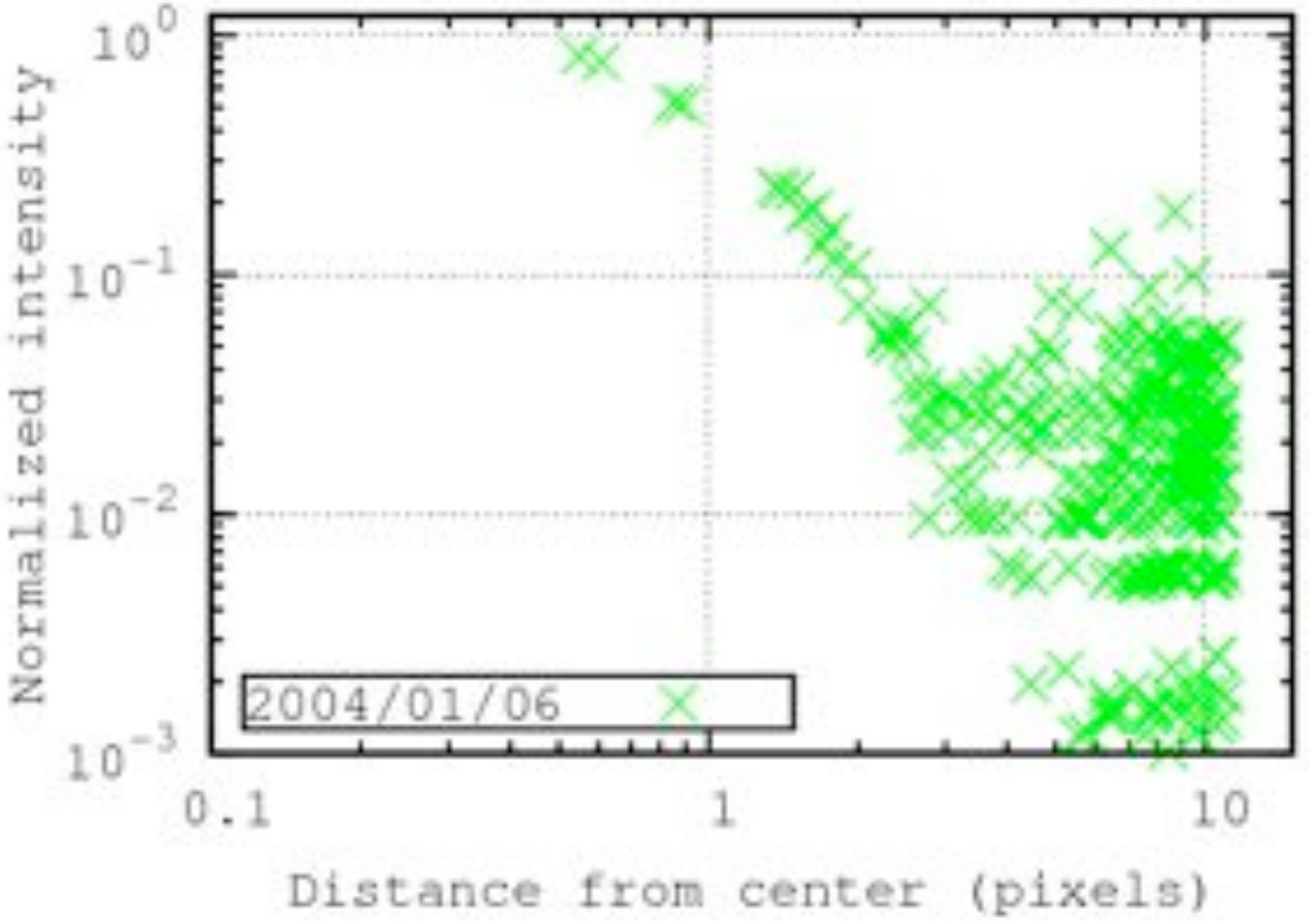

Figure 18.(a)

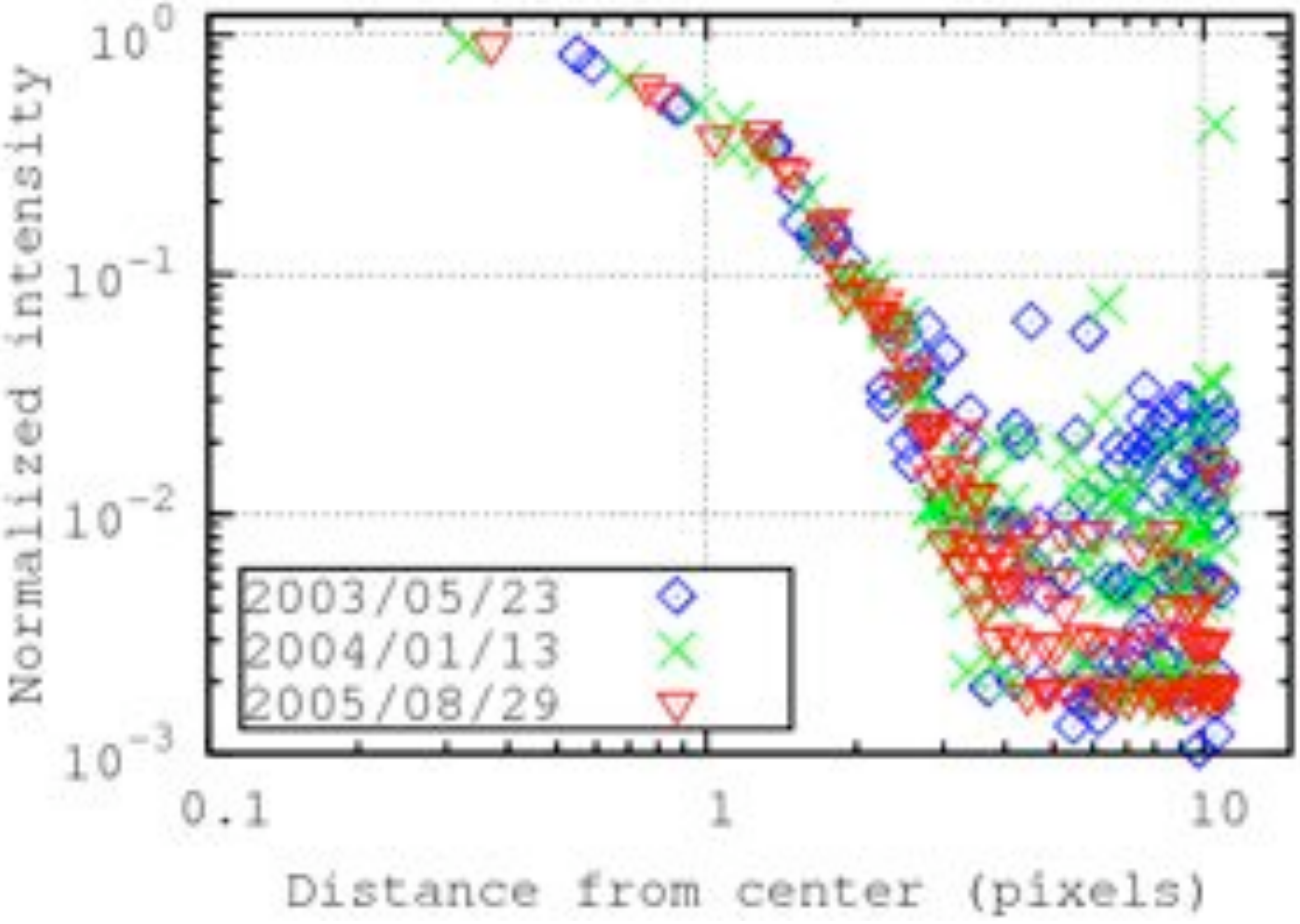

Figure 18. (b)

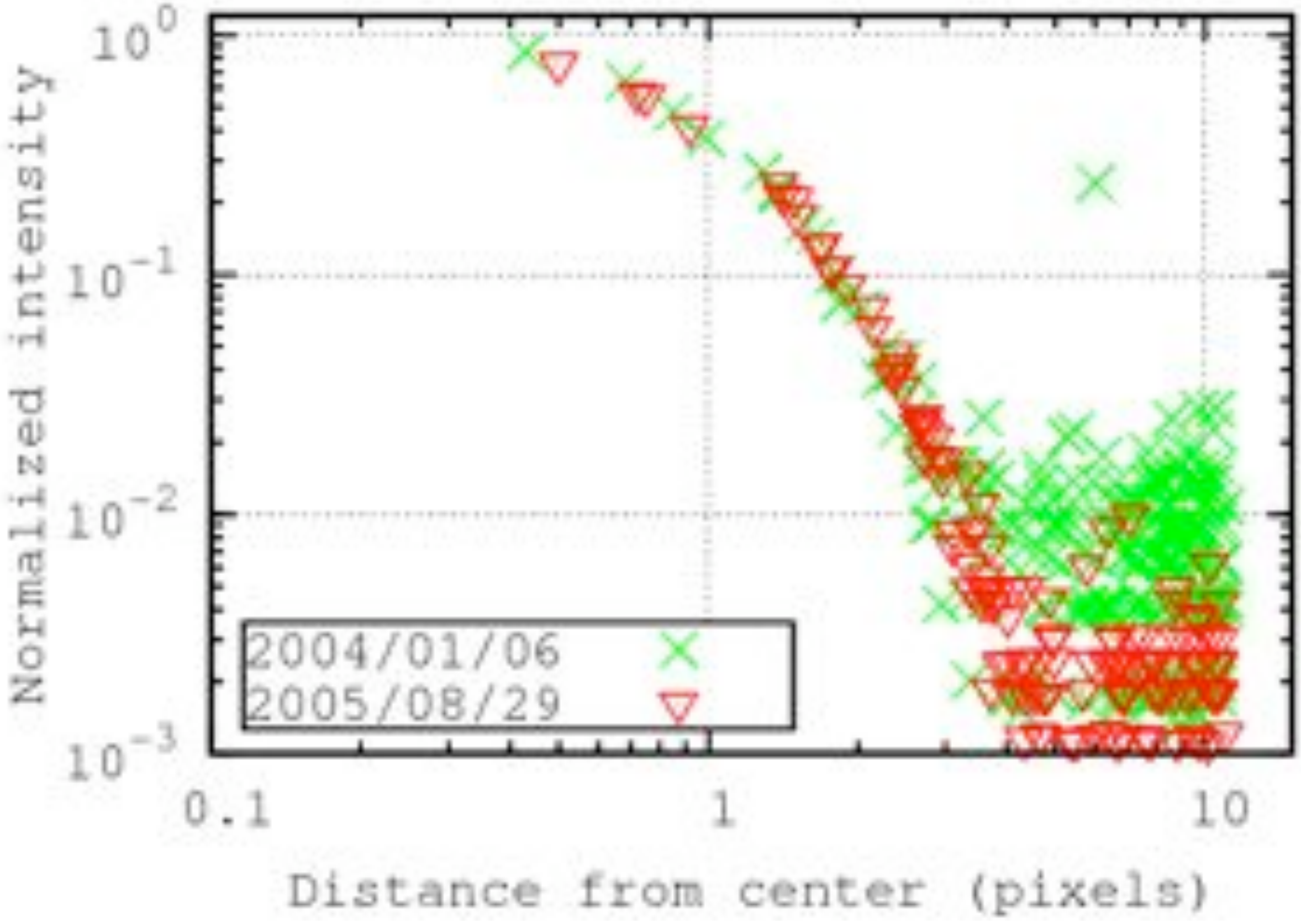

Figure 18.(c)

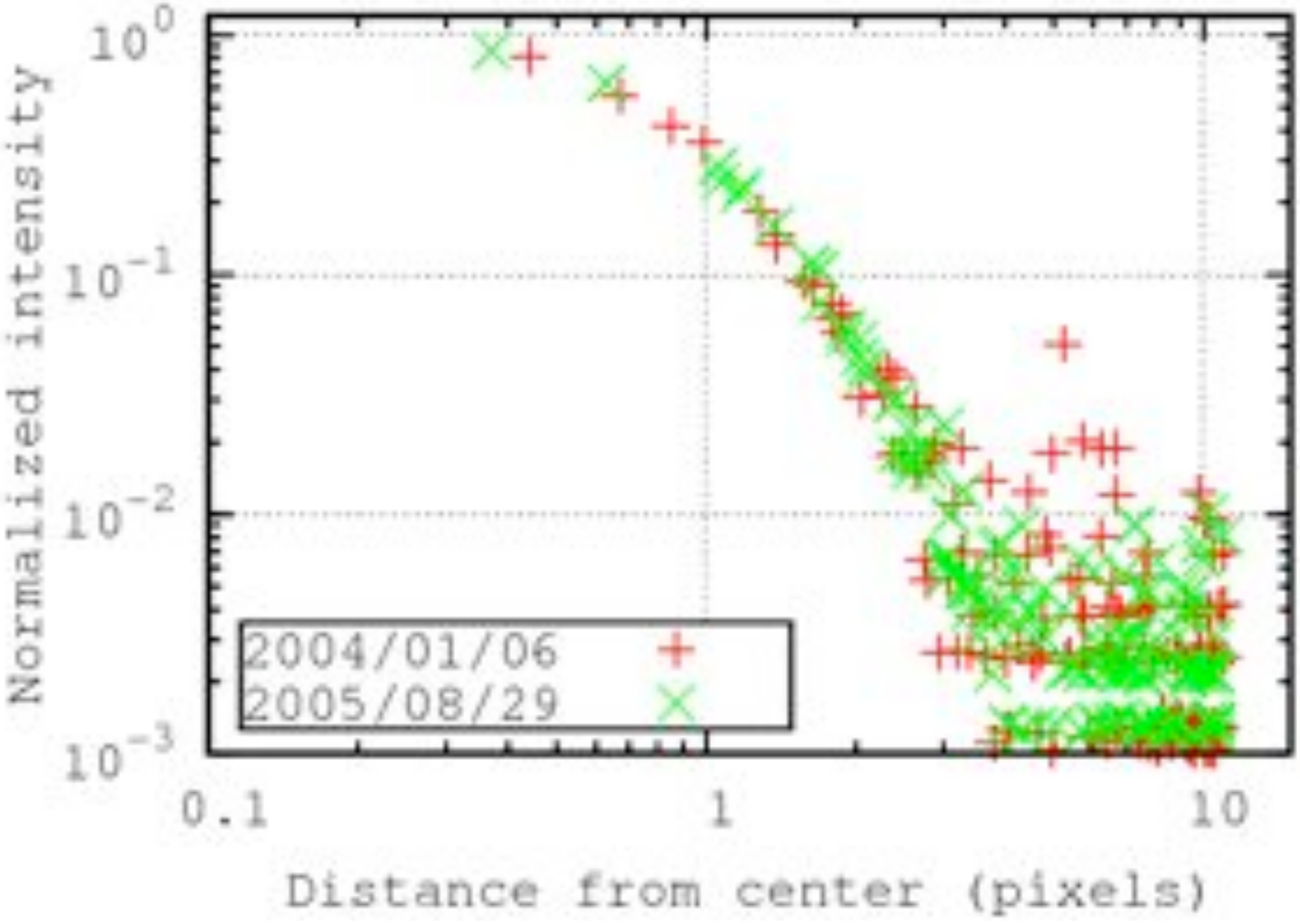

Figure 18.(d)

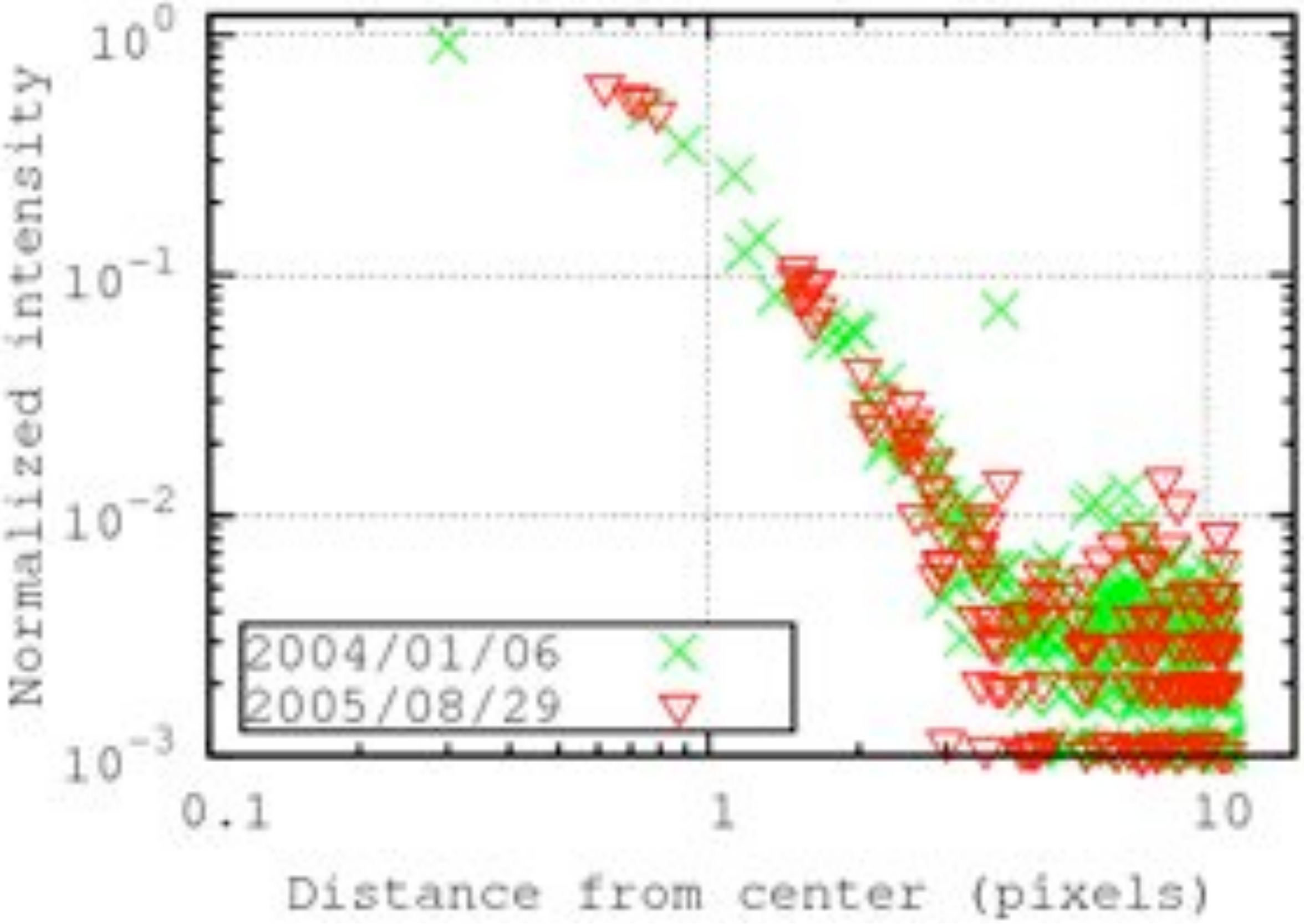

Figure 18. (e)

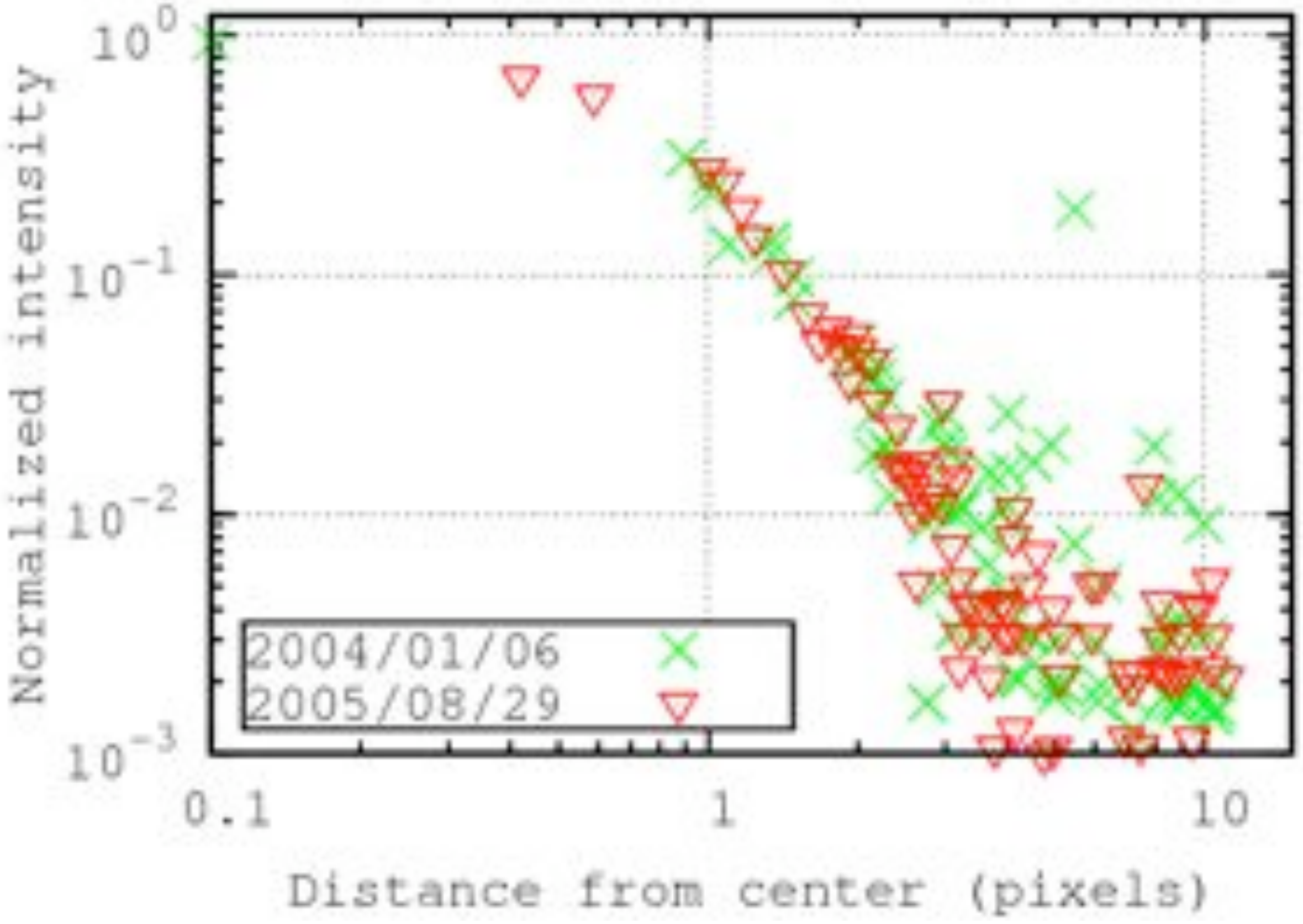

Figure 18.(f)

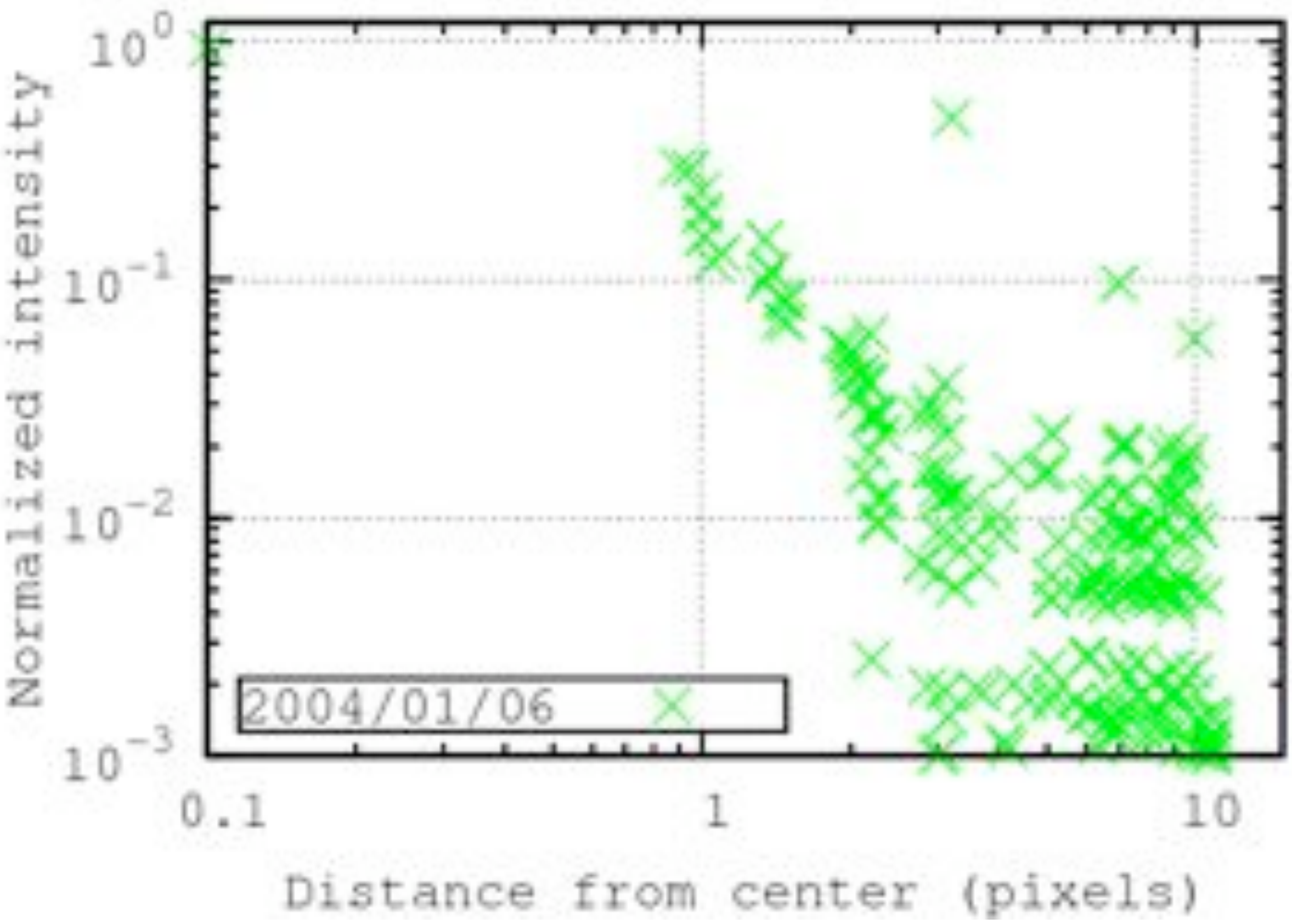

Figure 18.(g)

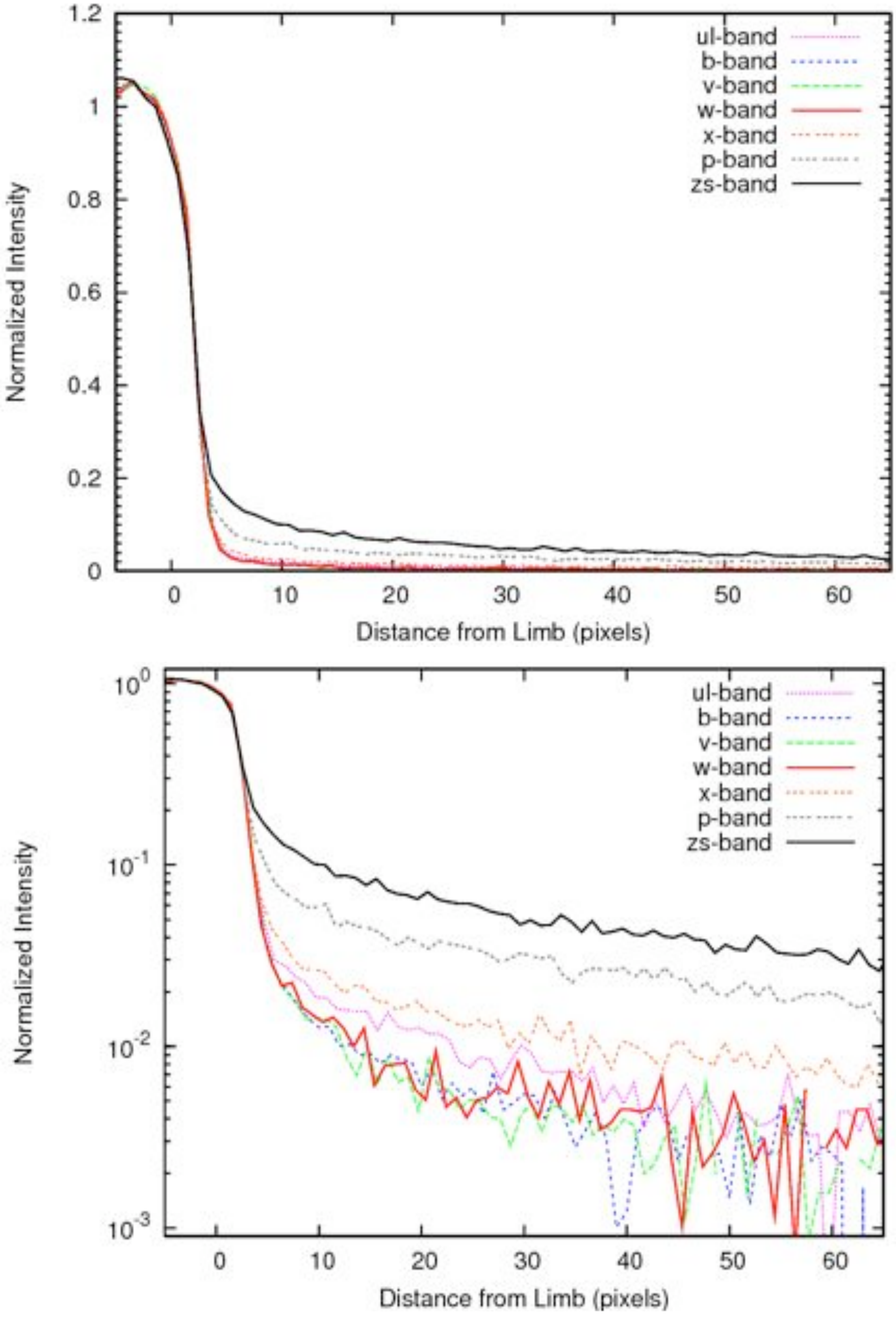

Figure 19.

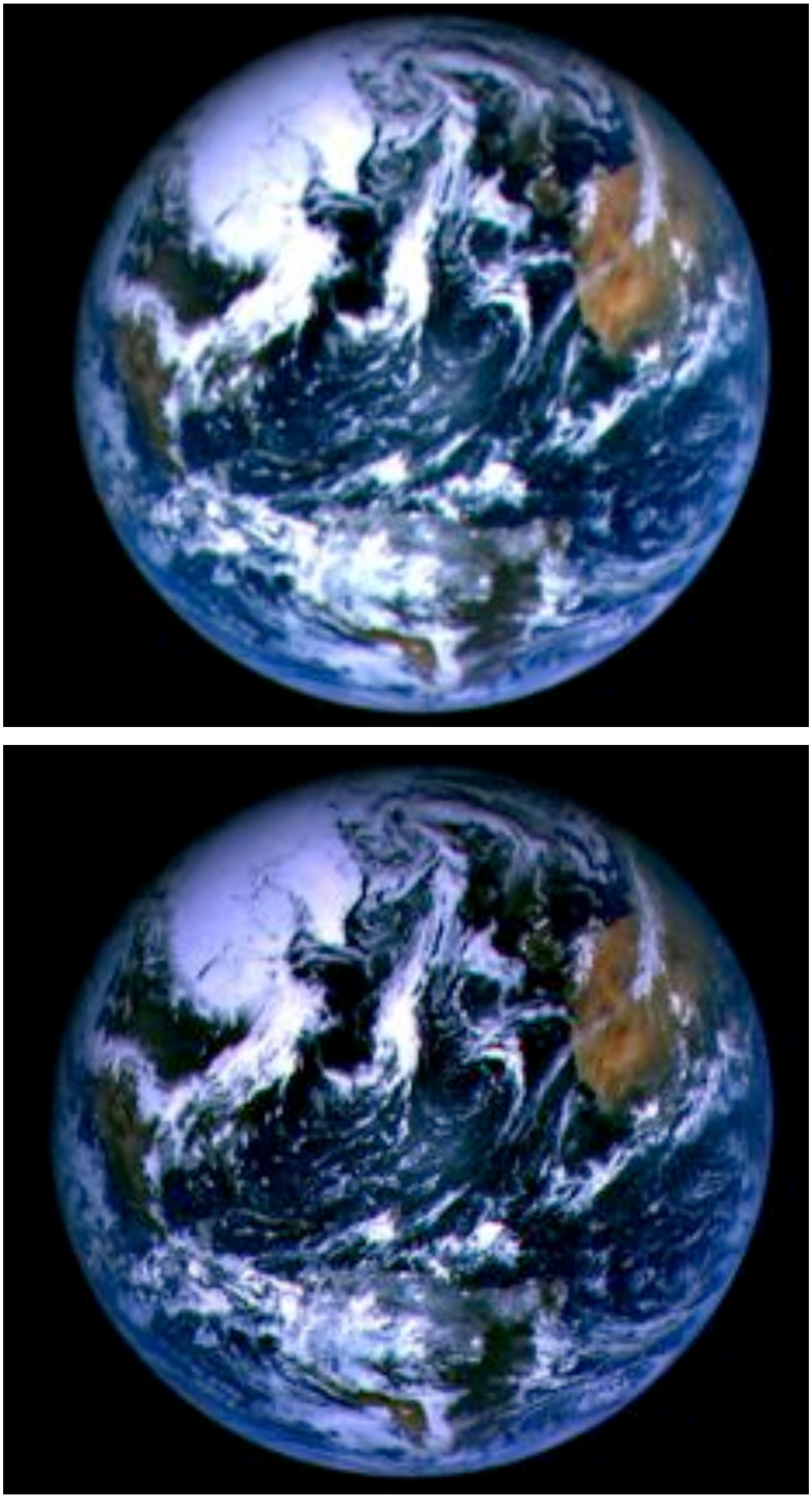

Figure 20.

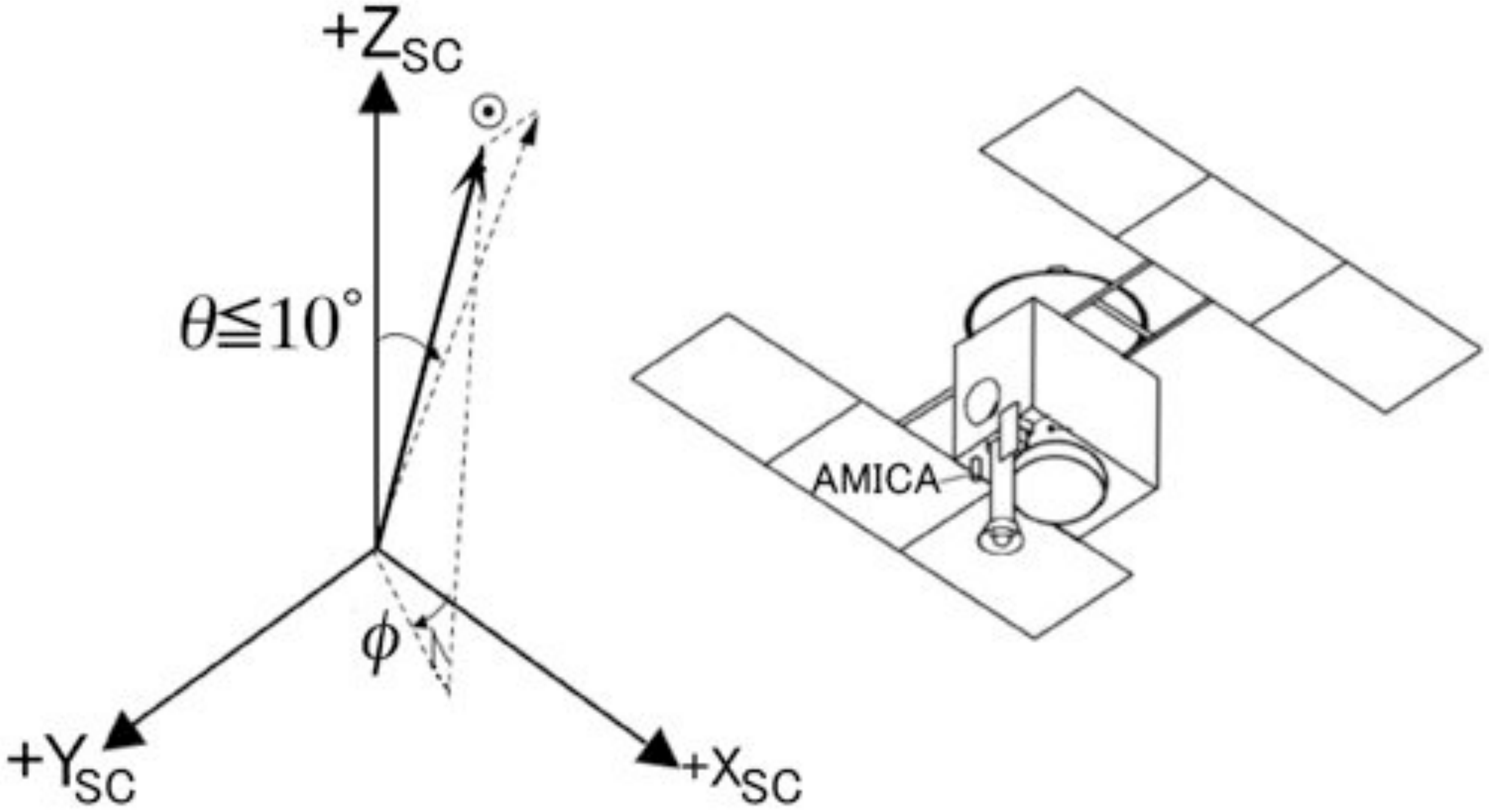

Figure 21

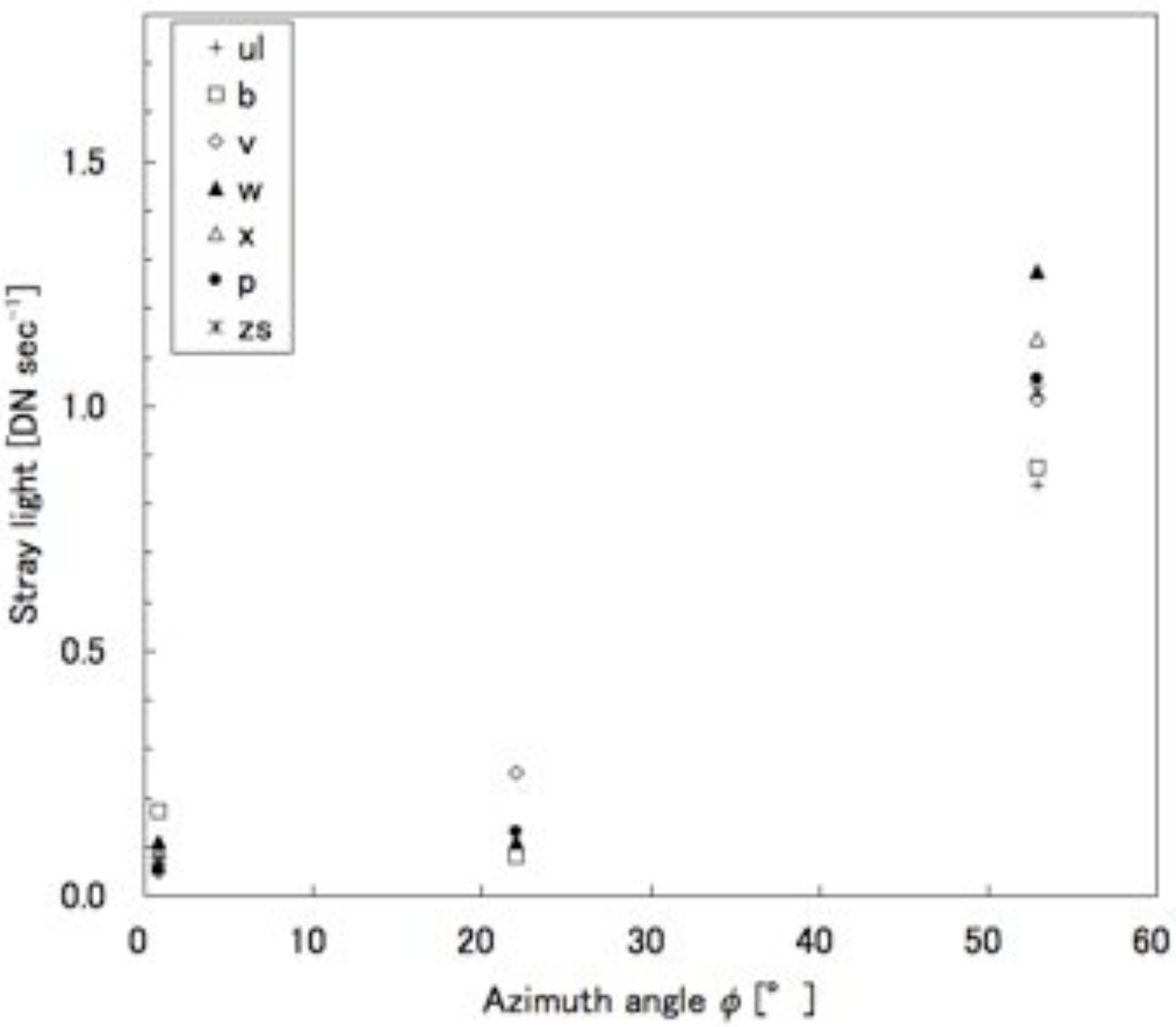

Figure 22.